# Phase transitions in two tunnel-coupled HgTe quantum wells. Bilayer graphene analogy and beyond

S. S. Krishtopenko[1,2], W. Knap[2], F. Teppe[2]*

[1]Institute for Physics of Microstructures RAS, GSP-105, 603950, Nizhni Novgorod, Russia

[2]Laboratoire Charles Coulomb (L2C), UMR CNRS 5221, Universite Montpellier, 34095 Montpellier, France

*Correspondence to: frederic.teppe@univ-montp2.fr

**Abstract**: HgTe quantum wells possess remarkable physical properties as for instance the quantum spin Hall state and the "single-valley" analog of graphene, depending on their layer thicknesses and barrier composition. However, double HgTe quantum wells yet contain more fascinating and still unrevealed features. Here we report on the study of the quantum phase transitions in tunnel-coupled HgTe layers separated by CdTe barrier. We demonstrate that this system has a 3/2 pseudo spin degree of freedom, which features a number of particular properties associated with the spin-dependent coupling between HgTe layers. We discover a specific metal phase arising in a wide range of HgTe and CdTe layer thicknesses, in which a gapless bulk and a pair of helical edge states coexist. This phase holds some properties of bilayer graphene such as an unconventional quantum Hall effect and an electrically-tunable band gap. In this "bilayer graphene" phase, electric field opens the band gap and drives the system into the quantum spin Hall state. Furthermore, we discover a new type of quantum phase transition arising from a mutual inversion between second electron- and hole-like subbands. This work paves the way towards novel materials based on multi-layered topological insulators.

**Main Text:**

Low energy band structure in graphene is formed by two spin-degenerate massless Dirac cones at two inequivalent valleys, giving rise to four massless Dirac cones in total (*1,2*). With one additional graphene layer added, bilayer graphene (BG) has an entirely different band structure. Most notably, symmetric BG is a zero-band gap semiconductor with quadratic energy-momentum dispersion (*3*). However, its band gap is continuously tunable through an electrical field applied perpendicularly to the sample (*4,5*). The electronic states in BG are also four-fold degenerate, taking into account both spin and valley degeneracies. Recently, it has been shown that HgTe quantum wells (QWs) with vanishing band gap possess a single spin-degenerate Dirac cone at the Brillouin zone center (*6-9*) and thus can be considered as the "single-valley" analog of graphene in two-dimensional (2D) semiconductor heterostructures.

The central feature of the HgTe QWs is the possibility of band inversion. The barrier material (e.g., CdTe) has a normal band ordering, with the *s*-type $\Gamma_6$ band lying above the *p*-type $\Gamma_8$ band, while in the well material (HgTe) the $\Gamma_6$ band lies below the $\Gamma_8$ band, corresponding to an inverted band ordering. If the thickness *d* of the HgTe/CdTe QWs does not exceed a critical value $d_c$, at the $\Gamma$ point, the first conduction subband (E1) is electron-like, i.e. it is formed by a linear combination of the $|\Gamma_6, m_J = \pm 1/2\rangle$ and $|\Gamma_8, m_J = \pm 1/2\rangle$ states, while the first valence subband (H1) arises from the $|\Gamma_8, m_J = \pm 3/2\rangle$ states, corresponding to the heavy-hole band. In wide HgTe QWs ($d > d_c$), the $E$1 subband falls below the $H$1 subband. The inversion between $E$1 and $H$1 subbands leads to the formation of a 2D time-reversal invariant topological insulator (TI) (*10*) denoted the quantum spin Hall (QSH) state (*6,7*) with dissipation less edge channel transport at zero external magnetic field (*11*). At critical QW thickness $d_c$, corresponding to topological phase transition between a TI and a trivial band insulator (BI), a low-energy band structure mimics a massless Dirac cone at the $\Gamma$ point (*6*).

Two tunnel-coupled HgTe QWs of critical thickness may, therefore, share some properties of natural BG. If each of the HgTe QW has $d > d_c$, this system offers a practical realization of tunnel-coupled layers of 2D TI. Moreover, the strained HgTe thick films have been proven to be three-dimensional (3D) TIs (*12,13*). Therefore, a wide double HgTe QW can be considered as two thick layers of 3D TI separated by BI (CdTe material) with the surface states arising at the interfaces (*14*). A tensile strain in the HgTe films, which opens a topological gap, is induced by the difference in the lattice parameters of HgTe and CdTe. Thus, double HgTe QWs is a realistic system, which potentially paves the way towards physics of multi-layered topological insulator materials.

So far, there are just a few works devoted to double HgTe QWs (*15-17*). All these works are based on the approach, in which each QW is described within Bernevig-Hughes-Zhang (BHZ) model (*6*) coupled by a spin-conserved tunneling Hamiltonian. The latter will be shown to be irrelevant on double HgTe QWs, which feature a much richer physics than previously assumed.

In our theoretical investigation of double HgTe QWs (see Fig. 1), we first start from band structure calculations on the basis of an eight-band Kane model (*18*). By using realistic material parameters, we obtain the phase diagram for symmetrical double QW as a function of the layer thicknesses. We discover a specific metal phase, in which the band properties in perpendicular magnetic and electric fields are very similar to the ones of natural BG. Then, we deduce an effective 2D Hamiltonian, involving two electron-like ($E$1, $E$2) and two hole-like ($H$1, $H$2) subbands, to describe quantum phase transitions in the system. On the basis of such a simplified Hamiltonian, we calculate dispersion of edge states in different phases. We show that inversion between $E$2 and $H$2 subbands induces an additional pair of helical edge states, putting the system into BI phase even in the case of inverted band structure. The crossing between $E$2 and $H$2 levels yields a Dirac cone in the $\Gamma$ point of the Brillouin zone.

Considering double HgTe QW, shown in Fig. 1, as a whole system and by using expansion in the plane-wave basis set (*19*), we calculate energy dispersions of electronic subbands. Figures 2A and 2B show positions of electron-like and hole-like subbands at $\boldsymbol{k} = 0$ as a function of middle barrier thickness $t$. Intuitively, it is clear that at infinitely large barrier, calculated states correspond to the subband positions in two separated HgTe QWs, while at finite values of $t$ each pair of the subbands is connected with even-odd state splitting in double QWs. If $t$ tends to zero, the energy values transform into positions of electronic subbands in single HgTe QWs of $2d$ thickness. It is seen that the splitting between electron-like levels exceeds significantly the one of the hole-like levels. This is due to significant difference in effective masses at $\boldsymbol{k} = 0$ in electron-like and hole-like subbands, which values determine the tunnel-coupling between the states and their even-odd splitting at given $t$.

There are two types of the band structure ordering, which arises depending on the HgTe layers thickness $d$. The first case shown in Fig. 2A takes place if $d$ varies in the range from $d_c/2$ to $d_c$, where $d_c$ corresponds to the thickness of single QW with Dirac cone at the Γ point. In this case $E$2 subband always lies above $H$2 subband and inversion between $E$1 and $H$1 subband can take place. At higher values of $t$, the $E$1 subband gains a higher energy than the $H$1 subband, and the system has the normal band structure. When the thickness $t$ is decreased, the energy of the $E$1 subband reduces, whereas the $H$1 subband energy practically does not change. The different dependence of $E$1 and $H$1 subbands on $t$ implies that at some barrier thicknesses the band gap closes. In fact, the crossing point between $E$1 and $H$1 subbands yields a low-energy band structure with Dirac cone in the vicinity of $k$ = 0 (*19*). If the values of $t$ are small enough to induce a gap between $H$1 and $H$2 subbands, it implies a quantum phase transition between BI and TI states, as in single HgTe QW (*6*). However, if $H$1 and $H$2 subbands coincide at $k$ = 0, it gives rise to an additional "massive" branch of valence subband at the crossing point, similar to the pseudospin-1 Dirac-Weyl 2D systems (*20*), as well as a quantum phase transition into a specific metal phase with the following subband ordering: $E$2-$H$1-$H$2-$E$1. The reasons to assign this metal phase as a BG phase are discussed later.

The second type of band structure ordering, shown in Fig. 2B, is realized when $d > d_c$. In this case $E$1 subband always lies below $H$1 subband, i.e. they are inverted. At lower barrier thickness, $E$2 subband lies above $H$2 subband, while at high values of $t$, they swap their positions. At specific values of $t$, $E$2 and $H$2 subbands cross, which also leads to the appearance of a Dirac cone at the Γ point. As for the case shown in Fig. 2A, additional "massive" branch arises as well. However, such branch corresponds to conduction $H$1 subband. Up to date, it was never realized that crossing between $E$2 and $H$2 subbands also induces a quantum phase transition. In this work we explicitly demonstrate it for the first time. In particular, we show that this crossing point in double HgTe QWs correspond to the quantum phase transition between BG and BI phase.

Figure 2C shows the phase diagram, in which two bold lines correspond to the Dirac cones at the Γ point. The left one results from the crossing between $E$1 and $H$1 subbands, while the right-side curve is connected with the crossing of $E$2 and $H$2 levels. If the middle barrier is thin enough, a gap between $H$1 and $H$2 subbands opens, and the inversion between $E$1 and H1 levels induces a quantum phase transition between BI and TI phases. The latter is shown by orange region. For relatively wide QWs, the so-called semimetal (SM) phase, corresponding to the white-striped region, is implemented. It is characterized by a vanishing indirect band gap, when the side maximum of the valence subband exceeds in energy the conduction subband bottom. This phase arises in single HgTe QWs as well when the thickness goes beyond some critical value, denoted in Fig. 2C by $d_{SM}$ (*21,22*). BG metal phase mentioned above corresponds to the blue region in Fig. 2C. Dispersion curves at various values of $d$ and $t$ in the vicinity of phase transitions between BI, BG and SM phases are provided in (*19*).

Let us now explain the reasons to call this specific metal phase with the ordering of electronic subbands $E$2-$H$1-$H$2-$E$1 a BG phase. First, we consider in details the case when both QWs have a critical thickness $d_c \approx 6.5$ nm, which at infinitely large barrier corresponds to two Dirac cones. The presence of the transparent barrier, for example of $t$ = 3 nm, turns from two Dirac cones in the vicinity of the Γ point into a band structure very similar to the one of natural BG (*3*) (c.f. Fig. 3A). In particular, it consists of two gapless isotropic parabolas, formed by $H$1 and $H$2 subbands. Moreover, as in BG in which non-zero band gap can be induced by breaking the inversion symmetry of two monolayers, in double HgTe QWs it could be obtained by using QWs of different thickness or by adding of one-side chemically doping. Moreover, the potential of a continuously tunable band gap through an electric field applied perpendicularly to the sample plane is of particular importance.

The double HgTe QW in BG phase also holds this property. Fig. 3B displays energy dispersion for BG phase in perpendicular electric field of 20 kV/cm. Even in this case the dispersion curves are very similar to the band structure of natural BG in external electric field (*3,4*). However, strong spin-orbit interaction in HgTe layers removes the spin degeneracy away from the Γ point due to Rashba effect (*23*). Fig. 3C shows the band gap values in the double QWs in BG phase as a function of applied electric field. Indeed, the band gap in BG phase is electrically-tunable, as it is in natural BG. However, its dependence on electric field has non-monotonic behaviour in double HgTe QWs. The reason for this is related with the additional side maximum of valence subband, which is also responsible for the formation of SM phase in zero electric field (see the diagram, shown in Fig. 2C). The top of the side maximum is increasing with the strength of electric field, and the band gap reduces (see the inset, shown in Fig. 3C). As a result, the band gap is closed in high enough electric field, giving rise to the formation of the SM phase. A critical electric field, corresponding to arising of indirect-band gap, coincides with the maximum value of Δ, shown in Fig. 3C.

Another characteristic of natural BG is the unconventional quantum Hall effect, related with the absence of zero-Landau level (LL) plateaus in Hall conductivity (*24*). For natural BG, plateaus in the Hall conductivity $\sigma_{xy}$, occur at integer multiples of $4e^2/h$. This is similar to a conventional semiconductor with level degeneracy $g = 4$ arising from the spin and valley degrees of freedom. Deviation from the conventional case occurs at low density, where there is a step in $\sigma_{xy}$ of height $8e^2/h$ across zero density, arising from the eightfold degeneracy of the zero-energy LL. This specific LL is formed by atomic orbitals of different sublattice sites from both layers (*25*).

Figure 3D shows LL fan chart in double HgTe QW with BG phase, calculated within the eight-band Kane model. In the BG phase, in perpendicular magnetic field a specific zero-mode LL with degeneracy two times higher than other LLs also arises. This zero-mode LL is shown in Fig. 3D, by bold orange line. For double HgTe QW, plateaus in the Hall conductivity are expected to occur at integer numbers of $e^2/h$. However, the doubled degeneracy of zero-mode LL requires twice as many carriers to fill them, so the transition between the corresponding plateaus should be twice as wide in density, and the step in $\sigma_{xy}$ between the plateaus are expected to be twice as high, $2e^2/h$ instead of $e^2/h$. Since the zero-mode LL in double HgTe QW is formed by the states of both *H*1 and *H*2 subbands, the presence of inversion asymmetry of two HgTe layers not only opens the band gap between *H*1 and *H*2 subbands (see Fig. 3B) but also splits the zero-mode LLs, removing the double degeneracy order. Therefore, we also expect recovering of sequence of equidistant plateaus in the Hall conductivity as it is for gate-biased natural BG (*5*).

Besides the zero-mode LL, there are two additional specific LLs, which in high magnetic fields are formed only by the states from *E*1 and *E*2 subbands. These LLs are shown in Fig. 3D by blue curves. In moderate magnetic fields, these *E*1 and *E*2 LLs are mixed with the states from *H*2 and *H*1 subbands respectively. It results in anticrossing between *E*1 (*E*2) LL and LLs from *H*2 (*H*1) subband, which is clearly seen in magnetic fields below 1 T. The crossing between the *E*1 and zero-mode LL, arising at critical magnetic field $B_c \sim 3.5$ T, corresponds to the transition from inverted into normal band structure, similar to this observed in single HgTe QW (*7*).

We have considered in details the case of double HgTe QW, when both HgTe layers have a critical thickness $d_c$. However, all the mentioned properties of BG phase hold for any double QW with the values of $d$ and $t$, corresponding to the blue region in Fig. 2C. The only difference is the ratio $M_1$ over $M_2$, where $M_1$ parameter describes the energy gap between $E1$ and $H1$ subbands, while $M_2$ corresponds to the half of the gap between $E2$ and $H2$ levels. If $d < d_c$, $2M_2$ exceeds $2M_1$, while in the opposite case of $d > d_c$, the gap between $E2$ and $H2$ subbands is lower than $2M_1$. An amazing property of double HgTe QW is that it shares some characteristics of natural BG even in the BI phase at $d > d_c$. In particular, the possibility of tuning the band gap by electric field and of observing the unconventional double step in plateaus in the Hall conductivity still persists.

We now discuss quantum phase transitions in double HgTe QWs. For this purpose, we derive an effective 2D Hamiltonian, which describes the band structure in the vicinity of $\mathbf{k} = 0$ in the phases shown in Fig. 2C. To infer this simplified model, we start from eight Bloch basic states, combined into an eight-component spinor:

$$\Psi = \left(|\Gamma_6, \tfrac{1}{2}\rangle, |\Gamma_6, -\tfrac{1}{2}\rangle, |\Gamma_8, \tfrac{3}{2}\rangle, |\Gamma_8, \tfrac{1}{2}\rangle, |\Gamma_8, -\tfrac{1}{2}\rangle, |\Gamma_8, -\tfrac{3}{2}\rangle, |\Gamma_7, \tfrac{1}{2}\rangle, |\Gamma_7, -\tfrac{1}{2}\rangle\right). \tag{1}$$

For the QWs grown in [001] direction, projection $m_J$ of total angular momentum at $\mathbf{k} = 0$ on the growth direction is still a good quantum number. At the $\Gamma$ point all QW subband states are formed by linear combination of the mentioned eight bulk bands. To describe BI, TI and BG phases, one should consider $E1$, $E2$, $H1$ and $H2$ subbands (see Fig. 2A and 2B). At $\mathbf{k} = 0$, the $|E1,\pm\rangle$ and $|E2,\pm\rangle$ subband states are formed from the linear combination of the $|\Gamma_6, m_J = \pm 1/2\rangle$, $|\Gamma_7, m_J = \pm 1/2\rangle$ and $|\Gamma_8, m_J = \pm 1/2\rangle$, while the $|H1,\pm\rangle$ and $|H2,\pm\rangle$ QW states are formed from the $|\Gamma_8, m_J = \pm 3/2\rangle$ states. Away from the $\Gamma$ point, the $E1$, $E2$, $H1$ and $H2$ states are mixed. To construct the effective Hamiltonian one should carefully take into account different parities of the envelope function components from $|\Gamma_6, \pm 1/2\rangle$, $|\Gamma_7, \pm 1/2\rangle$, $|\Gamma_8, \pm 1/2\rangle$ and $|\Gamma_8, \pm 3/2\rangle$ bulk bands into formation of given subband state (*26*). For instance, $E1$ state is formed by even function corresponding to $|\Gamma_6, \pm 1/2\rangle$ band and by odd envelope functions corresponding to $|\Gamma_7, \pm 1/2\rangle$ and $|\Gamma_8, \pm 1/2\rangle$ bands, while parities of their contributions into $E2$ state are changed. In previous works (*15-17*), this point was missed; it results in the wrong description of double HgTe QWs. Detailed explanation is given in (*19*).

After straightforward calculations, described in (*19*), we obtain the following form of the effective 2D Hamiltonian for the $E1$, $E2$, $H1$, $H2$ states, expressed in the basis of two Kramer's sets $|E1,+\rangle$, $|H1,+\rangle$, $|H2,-\rangle$, $|E2,-\rangle$ and $|E2,+\rangle$, $|H2,+\rangle$, $|H1,-\rangle$, $|E1,-\rangle$:

$$H_{eff}(k_x, k_y) = \begin{pmatrix} H(k_x, k_y) & 0 \\ 0 & \Theta H(k_x, k_y) \Theta^{-1} \end{pmatrix}, \tag{1}$$

where $\Theta$ is a "time reversal" operator, given by

$$\Theta = \begin{pmatrix} 0 & i\sigma_y \\ i\sigma_y & 0 \end{pmatrix} K, \tag{2}$$

with $K$ stands for complex conjugation and $\sigma_y$ is one of the Pauli spin matrices. Each block in (1) is described by a four-component spinor with pseudospin $J = 3/2$ degree of freedom (*19*). $H(k_x, k_y)$ is written as

$$H(k_x, k_y) = \begin{pmatrix} \varepsilon_{E1}(k) & -A_1 k_+ & R_1 k_-^2 & S_0 k_- \\ -A_1 k_- & \varepsilon_{H1}(k) & 0 & R_2 k_-^2 \\ R_1 k_+^2 & 0 & \varepsilon_{H2}(k) & A_2 k_+ \\ S_0 k_+ & R_2 k_+^2 & A_2 k_- & \varepsilon_{E2}(k) \end{pmatrix},$$

$$k_\pm = k_x + ik_y,$$

$$\varepsilon_{E1}(k) = C + \frac{\Delta_{H1H2}}{2} + 2M_1 + B_{E1}(k_x^2 + k_y^2),$$

$$\varepsilon_{H1}(k) = C + \frac{\Delta_{H1H2}}{2} + B_{H1}(k_x^2 + k_y^2),$$

$$\varepsilon_{H2}(k) = C - \frac{\Delta_{H1H2}}{2} + B_{H2}(k_x^2 + k_y^2),$$

$$\varepsilon_{E2}(k) = C - \frac{\Delta_{H1H2}}{2} + 2M_2 + B_{E1}(k_x^2 + k_y^2). \tag{3}$$

Here, $k_x$ and $k_y$ are momentum components in the plane of double QW, and $C$, $M_1$, $M_2$, $A_1$, $A_2$, $B_{E1}$, $B_{H1}$, $B_{H2}$, $B_{E2}$, $\Delta_{H1H2}$, $R_1$, $R_2$, $S_0$ are specific heterostructure constants, being defined by QW geometry and materials. The Hamiltonian $H_{eff}(k_x, k_y)$ has block-diagonal form because we keep the inversion symmetry and axial symmetry around the growth direction (*19*). We note that $H_{eff}(k_x, k_y)$ is valid for any values of $d$ and $t$. In particular for $t = 0$, it describes energy dispersion in the vicinity of the Γ point in single HgTe QW beyond the BHZ model. Parameters $\Delta_{H1H2}$, $R_1$, $R_2$, $S_0$ significantly depend on $t$, and all tend to zero at large middle barrier thickness, while $B_{E1} = B_{E2}$ coincides. As it is easy to see, in this case the system is described by two non-interacting BHZ models, written for two pairs formed by $E1$, $H1$ and $E2$, $H2$ subbands. We note that $B_{H1} = B_{H2}$ if $\Delta_{H1H2} = 0$.

The most important quantities in $H(k_x, k_y)$ are two mass parameters $M_1$ and $M_2$. Almost all phases in Fig. 2C can be grouped into three types according to the sign of $M_1$ and $M_2$. The BI phase at normal band structure, shown in the left part of the diagram, corresponds to positive values of $M_1$ and $M_2$. The BG and TI phases arise if $M_1 < 0$ and $M_2 > 0$, and the difference between these phases is connected with $\Delta_{H1H2}$, which equals to zero in the case of BG phase. The second BI phase, shown in the right from the BG phase, corresponds to negative values of $M_1$ and $M_2$, and $\Delta_{H1H2} = 0$. The SM phase, arising at all values of $M_2$ and $M_1 > 0$, cannot be described within the simplified model because it is only valid near the Γ point, while the SM phase is formed by non-local overlapping of the valence band and conduction band bottom (*21,22*). Comparison between calculations of subband curves, performed within the simplified model and on the basic of an eight-band Kane Hamiltonian for different phases is given in (*19*).

Zero values of $M_1$ and $M_2$ both conform to quantum phase transitions at which Dirac cone at the Γ point occurs. The left and right bold black curves in the phase diagram, shown in Fig. 2C, are related with $M_1 = 0$ and $M_2 = 0$ respectively. Figure 4 presents energy dispersion for the bulk and edge states in various phases in double HgTe QWs obtained within the simplified model. Dispersion of the bulk states are shown in black, while the orange and blue curves are the edge states, described by different blocks of $H_{eff}(k_x, k_y)$. The electrons in the edge states, marked by different colors, move in opposite directions. Figure 4 demonstrates that each change of sign of mass parameter $M_1$ or $M_2$ yields a pair of helical edge states, providing a quantum phase transition at $M_1 = 0$ or $M_2 = 0$. In particular, BG phase arising at $M_1 < 0$ and $M_2 > 0$ has a single pair of the edge states, coexisting with the gapless bulk states. If $M_1 < 0$ and $M_2 = 0$, the double HgTe QW mimics a 2D system with the presence of both bulk and edge massless fermions, which energy dispersions are shown in Fig. 4C.

Figure 4D shows special case with inversion between $E2$ and $H2$ subbands, which was not considered in 2D systems so far. It perfectly illustrates that each crossing between electron-like and hole-like levels of higher indexes, also results in the appearance of a pair of helical edge states. Topological properties of corresponding insulator phase are determined by amount of inverted levels, which are connected with the number of pairs of the edge states. In the particular case of inversion of both $E1$ with $H1$ subband and $E2$ with $H2$ subband, the terms proportional to $R_1$, $R_2$ and $S_0$ induce the coupling between counter-propagating states with the same spin orientation. The latter could be interpreted as a spin-dependent tunneling between two layers of 2D TI (*19*). It is clear from the inset that such tunnel-coupling opens the gap in the energy spectrum of the edge states even without external one-particle scattering processes. Thus, we prove that double HgTe QW with $M_1 < 0$ and $M_2 < 0$ has a BI phase.

Let us now remark the main difference between BG phase in double HgTe QW and natural BG. The electrons in natural BG are chiral particles with $S = 1/2$ pseudospin of freedom (*24,25*) and pseudospin winding number of 2 (*26*). It results from the commutation of operator $\vec{k} \cdot \vec{S}$ with the low-energy Hamiltonian of natural BG. As it is shown above, the simplest Hamiltonian, required for the description of bulk and edge states in the BG phase, has $J = 3/2$ pseudospin of freedom. Moreover, it can be shown that operator $\vec{k} \cdot \vec{J}$ does not commute with $H(k_x, k_y)$ (*19*), proving the non-chiral character of electron in double HgTe QWs. Thus, even though BG phase mimics some characteristics of natural BG, it is a novel and fascinating state of matter, exhibiting the coexistence of gapless bulk states and spin-polarized edge channels.

Since we have demonstrated the existence of gapless spin-polarized counter-propagating edge channels in double HgTe QWs, they should exhibit QSH effect in both TI ($\Delta_{H1H2} \neq 0$) and BG ($\Delta_{H1H2} = 0$) phases. Intuitively it is clear that QSH effect in the TI phase could be observed on the sample with a six-terminal Hall bar, as it has been previously proposed for single QW (*6,7*). However, the measurements of two-terminal conductance in the BG phase contain contributions both from the bulk states and from the helical edge modes. To separate these contributions, we propose to introduce inversion asymmetry between the HgTe layers. If the band gap for the bulk states is open, for example by an external electric field, the system is driven to the TI regime, and the detection of the QSH effect becomes possible (*6*).

**Acknowledgments:** We thank all members of the group of D. Carpentier in ENS Lyon for insightful discussions and comments of this work. The authors gratefully acknowledge Sandra Ruffenach for her assistance in manuscript preparation. This work was supported by the CNRS through LIA TeraMIR project, by Languedoc-Roussillon region via the "Terapole Gepeto Platform", by the Russian Academy of Sciences, the non-profit Dynasty foundation, the Russian Foundation for Basic Research (Grants 15-02-08274, 15-52-16012, 16-02-00672) and by Russian Ministry of Education and Science (Grant Nos. MK-6830.2015.2).

**References and Notes:**

1. K. S. Novoselov *et al.*, Two-dimensional gas of massless Dirac fermions in graphene. *Nature (London)* **438**, 197 (2005).

2. Y. Zhang, Y.-W. Tan, H. L. Stormer, P. Kim, Experimental observation of the quantum Hall effect and Berry's phase in graphene. *Nature (London)* **438**, 201 (2005).

3. T. Ohta, A. Bostwick, T. Seyller, K. Horn, E. Rotenberg, Controlling the Electronic Structure of Bilayer Graphene. *Science* **313**, 951 (2006).

4. E. McCann, Asymmetry gap in the electronic band structure of bilayer graphene. *Phys. Rev. B* **74**, 161403 (2006).

5. E. V. Castro *et al.*, Biased Bilayer Graphene: Semiconductor with a Gap Tunable by the Electric Field Effect. *Phys. Rev. Lett.* **99**, 216802 (2007).

6. B. A. Bernevig, T. L. Hughes, S. C. Zhang, Quantum Spin Hall Effect and Topological Phase Transition in HgTe Quantum Wells. *Science* **314**, 1757 (2006).

7. M. König *et al.*, Quantum Spin Hall Insulator State in HgTe Quantum Wells. *Science* **318**, 766 (2007).

8. B. Büttner *et al.*, Single valley Dirac fermions in zero-gap HgTe quantum wells. *Nat. Phys.* **7**, 418 (2011).

9. M. Zholudev *et al.*, Magnetospectroscopy of two-dimensional HgTe-based topological insulators around the critical thickness. *Phys. Rev. B* **86**, 205420 (2012).

10. C. L. Kane, E. J. Mele, $Z_2$ topological order and the quantum spin Hall effect. *Phys. Rev. Lett.* **95**, 146802 (2005).

11. A. Roth *et al.*, Nonlocal transport in the quantum spin Hall state. *Science* **325**, 294 (2009).

12. X. Dai, *et al.* Helical edge and surface states in HgTe quantum wells and bulk insulators. *Phys. Rev. B* **77**, 125319 (2008).

13. C. Brüne, *et al.* Quantum Hall Effect from the Topological Surface States of Strained Bulk HgTe *Phys. Rev. Lett.* **106**, 126803 (2011).

14. M. I. Dyakonov A. V. Khaetskii, Surface states in a gapless semiconductor. *JETP Lett.* **33**, 110 (1981).

15. P. Michetti, J. C. Budich, E. G. Novik, P. Recher, Tunable quantum spin Hall effect in double quantum wells. *Phys. Rev. B* **85**, 125309 (2012).

16. P. Michetti, B. Trauzettel, Devices with electrically tunable topological insulating phases. *Appl. Phys. Lett.* **102**, 063503 (2013).

17. J. C. Budich, B. Trauzettel, P. Michetti, Time Reversal Symmetric Topological Exciton Condensate in Bilayer HgTe Quantum Wells. *Phys. Rev. Lett.* **112**, 146405 (2014).

18. E. G. Novik *et al.*, Band structure of semimagnetic $Hg_{1-y}Mn_yTe$ quantum wells. *Phys. Rev. B* **72**, 035321 (2005).

19. See *Supplementary materials*.

20. J. D. Malcolm, E. J. Nicol, Magneto-optics of general pseudospin-*s* two-dimensional Dirac-Weyl fermions. *Phys. Rev. B* **90**, 035405 (2014).

21. Z. D. Kvon, *et al.*, Two-dimensional electron-hole system in a HgTe-based quantum well. *JETP Letters* **87**, 502 (2008).

22. Z. D. Kvon, *et al.*, Two-dimensional electron-hole system in HgTe-based quantum wells with surface orientation (112). *Phys. Rev. B* **83**, 193304 (2011).

23. E. I. Rashba, Properties of semiconductors with an extremum loop. 1. Cyclotron and combinational resonance in a magnetic field perpendicular to the plane of the loop. *Sov. Phys. Solid. State* **2**, 1109 (1960).

24. K. S. Novoselov, *et al.*, Unconventional quantum Hall effect and Berry's phase of $2\pi$ in bilayer graphene. *Nature Physics* **2**, 177 (2006).

25. E. McCann, V. I. Fal'ko, Landau-Level Degeneracy and Quantum Hall Effect in a Graphite Bilayer. *Phys. Rev. Lett.* **96**, 086805 (2006).

26. C.-H. Park, N. Marzari, Berry phase and pseudospin winding number in bilayer graphene. *Phys. Rev. B* **84**, 205440 (2011).

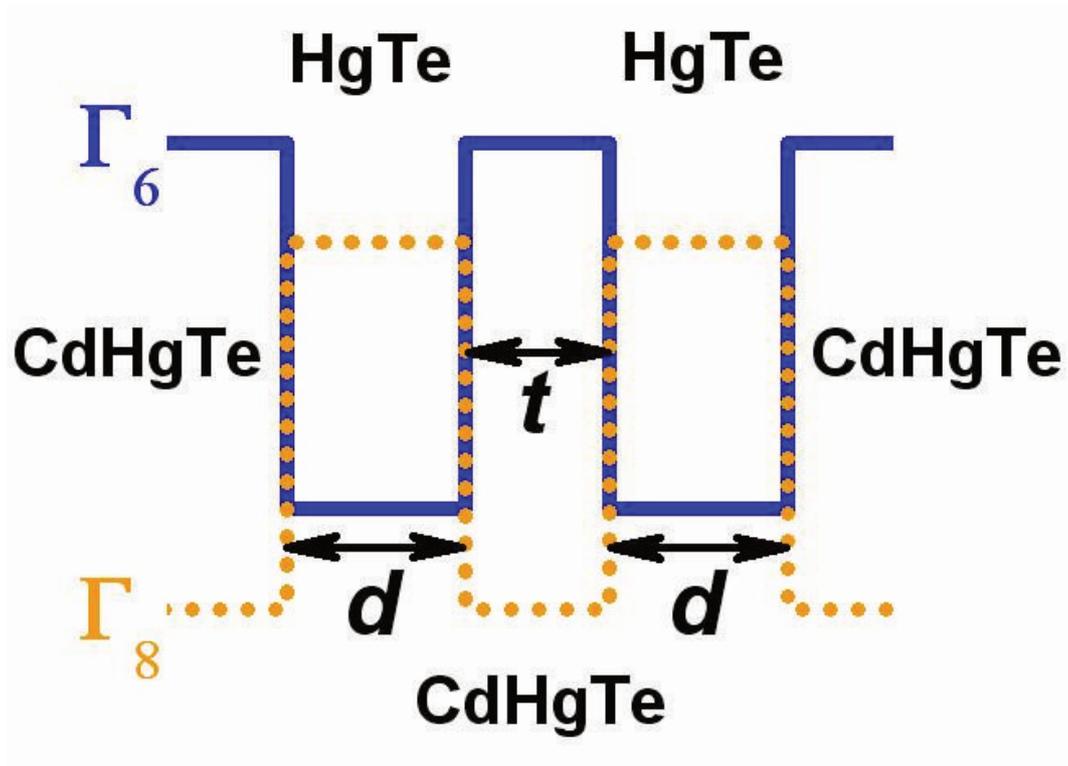

**Fig. 1**. Schematic representation of the double HgTe/CdHgTe QW. Here *d* is the thickness of HgTe layers and *t* is the middle barrier thickness. Further, we consider the double QW grown on CdTe buffer in (001) crystallographic direction. The concentration of mercury in the top, middle and lower barriers is assumed to be equal to 0.3 (*7,8*).

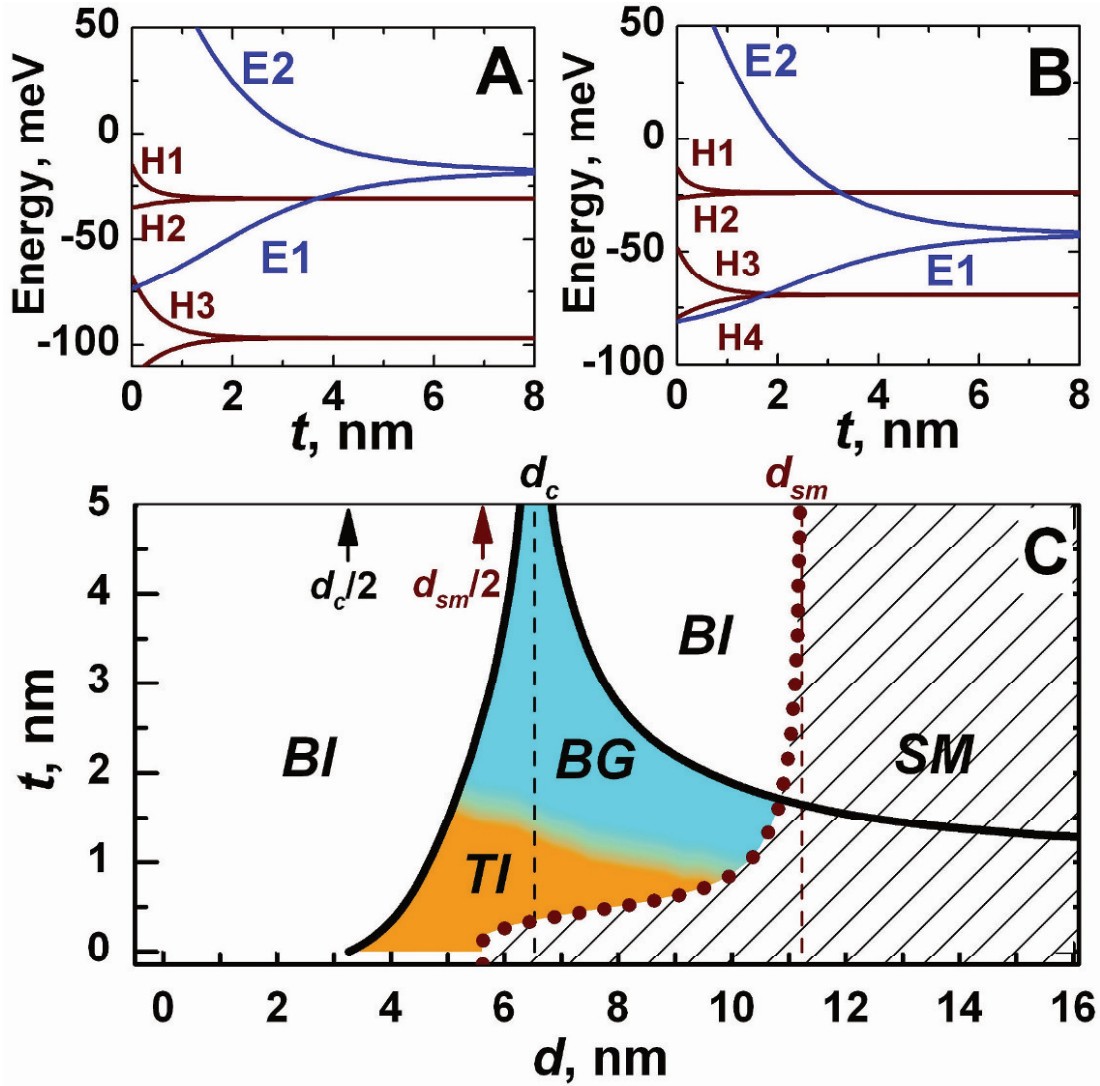

**Fig. 2**. (**A,B**) Energy of $E1$, $E2$ (both in blue) and $H1$, $H2$ (both in red) bands at $k = 0$ versus barrier thickness $t$ at different quantum well thickness $d$: (**A**) $d_c/2 < d < d_c$ and (**B**) $d > d_c$. (**C**) Phase diagram of double HgTe QW. The values $d_c$ and $d_{sm}$ correspond to thickness of the single QW, at which Dirac cone and semimetal phase arise respectively. The white-open regions are the band insulator phase, while the white-striped region corresponds to the semimetal phase, when the side maxima of valence subband exceed the bottom of conduction subband. The orange and blue regions conform to topological insulator and BG phase, respectively. The bold black curves correspond to the arising of the Dirac cone at the Γ point. We note that the scales of $d$ and $t$ in the phase diagram can be efficiently increased by changing $x$ and $y$ in the alloys of double $Hg_yCd_{1-y}Te/Cd_xHg_{1-x}Te$ QWs.

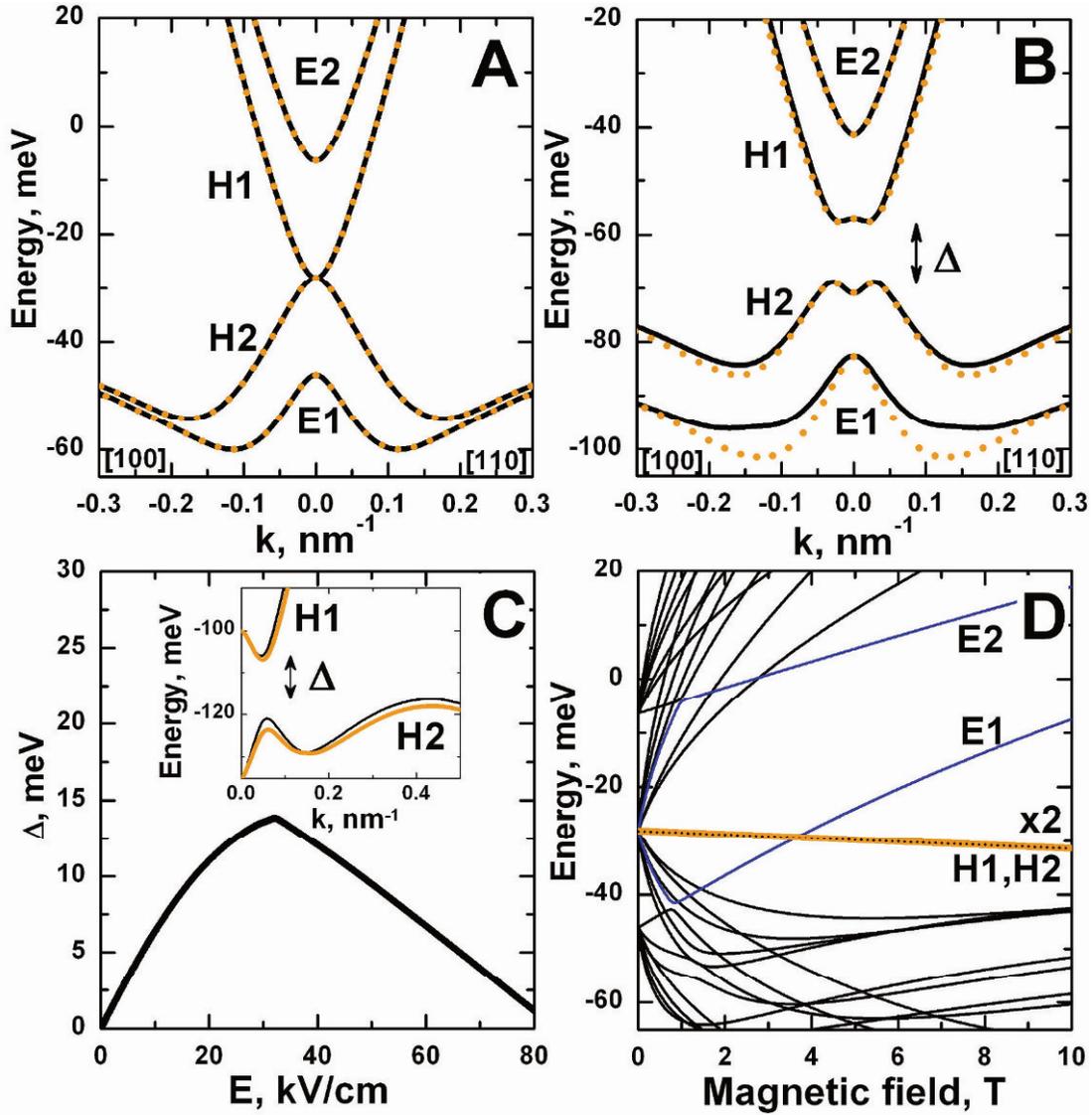

**Fig. 3**. (**A,B**) Energy dispersions for BG phase, implemented at $d = d_c \approx 6.5$ nm and $t = 3$ nm, in zero electric field (**A**) and in electric field of 20 kV/cm (**B**), oriented perpendicular the QW plane. Solid black and dotted orange curves correspond to different spin states. The presence of electric field not only opens the band gap $\Delta$ but also leads to the Rashba spin splitting (*23*). (**C**) Bulk band gap as a function of applied electric field. As it is in natural BG (*4,5*), double HgTe QW in this phase has also electrically-tunable band gap. However, the presence of additional side maximum in the valence subband closes the gap in high electric field. The inset shows energy dispersions in electric field of 50·kV/cm. Solid black and orange curves correspond to different spin states. (**D**) The Landau-level fan chart for BG phase. The zero-mode LL, which has doubled degeneracy order as compared with other levels, is marked by bold orange line. This LL is formed by states of both *H*1 and *H*2 subbands. LLs, containing only the states from *E*1 and *E*2 subbands in high magnetic fields, are given in blue. The crossing between the *E*1 and zero-mode LL, arising at critical magnetic field $B_c \sim 3.5$ T, leads to the phase transition into normal (non-inverted) band structure, as it is in single HgTe QW (*7*).

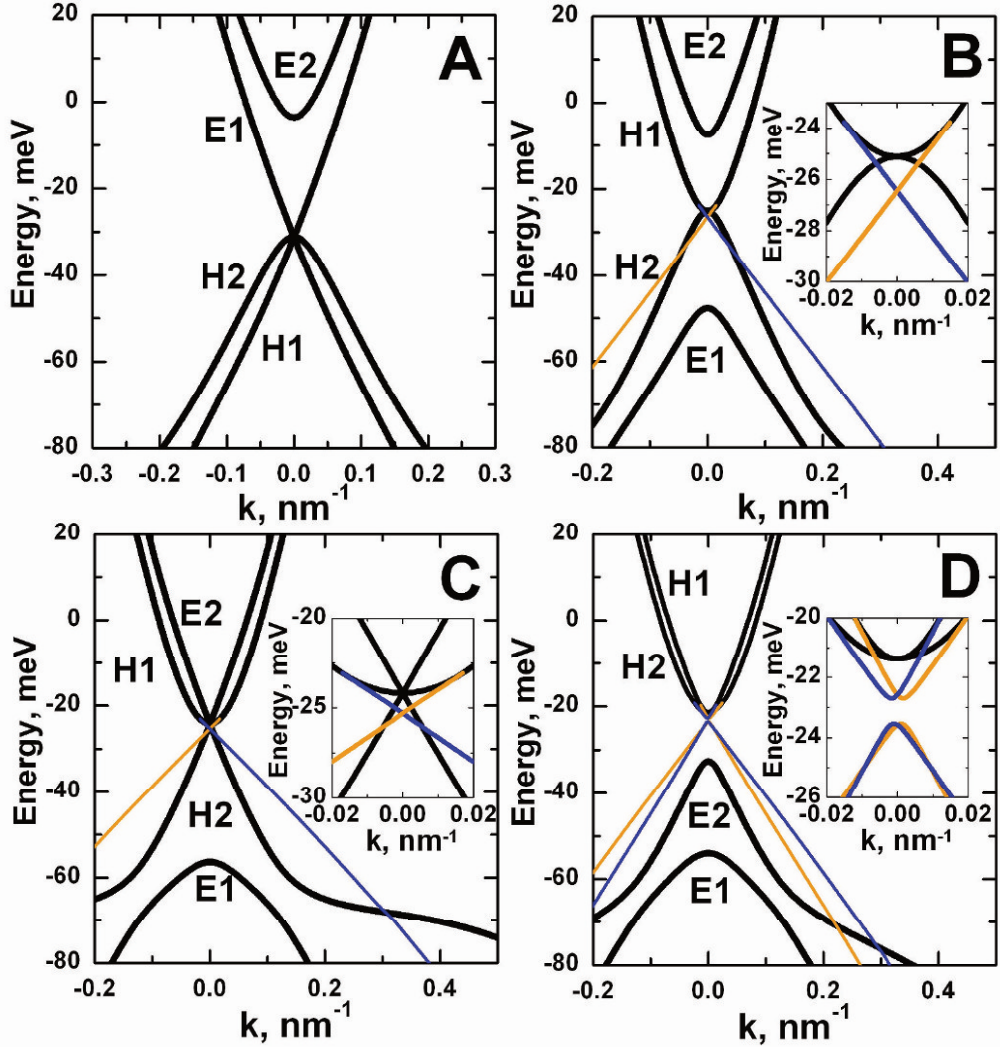

**Fig. 4.** Energy dispersions, calculated by using effective 2D Hamiltonian for (**A**) $M_1 = 0$ and $M_2 > 0$, (**B**) $M_1 < 0$ and $M_2 > 0$ (BG phase), (**C**) $M_1 < 0$ and $M_2 = 0$, (**D**) $M_1 < 0$ and $M_2 < 0$. Other band parameters are given in (*19*). Bulk states are shown in black. Orange and blue curves correspond to the dispersion of the edge states, obtained with open boundary conditions. Kramer's partners of the edge states, moving in opposite directions, are shown in different colors. The insets show the behavior of the dispersion curves in the vicinity of $k = 0$. It is seen from (**A**) and (**C**) that $M_1 = 0$ or $M_2 = 0$ corresponds to topological phase transition with arising of Dirac cone at the Γ point of the Brillouin zone, while each negative values of $M_1$ and $M_2$ results in appearance of additional pair of the helical edge states. The BG phase (**B**) with $M_1 < 0$ and $M_2 > 0$ is nontrivial and characterized by coexistence of gapless bulk states with single pair of the helical edge states. The case of $M_1 < 0$ and $M_2 < 0$ shown in (**D**) is characterized by the presence of two pairs of the edge states. Two edge states with the same spin couple to produce a gap in the spectrum, destroying the QSH effect and putting the system into trivial phase even with inverted band structure in double HgTe QW.

# Supplementary materials
## 8-band Kane model

We consider double HgTe/Cd$_{0.7}$Hg$_{0.3}$Te QW grown on (001)-oriented CdTe buffer and assume that the $z$ axis coincides with the crystallographic direction (001), while the $x$ and $y$ axes correspond to directions (100) and (010), respectively. To correctly account for the influence of nonparabolicity, spin-orbit interaction and lattice-mismatch deformation on electronic states in the double QW, we use the eight band Kane model (*17*). In the basis of Bloch amplitudes for the $\Gamma_6$, $\Gamma_8$ and $\Gamma_7$ bands

$$u_1(\mathbf{r}) = |\Gamma_6, +1/2\rangle = S\uparrow,$$

$$u_2(\mathbf{r}) = |\Gamma_6, -1/2\rangle = S\downarrow,$$

$$u_3(\mathbf{r}) = |\Gamma_8, +3/2\rangle = (1/\sqrt{2})(X + iY)\uparrow,$$

$$u_4(\mathbf{r}) = |\Gamma_8, +1/2\rangle = (1/\sqrt{6})[(X + iY)\downarrow - 2Z\uparrow],$$

$$u_5(\mathbf{r}) = |\Gamma_8, -1/2\rangle = -(1/\sqrt{6})[(X - iY)\uparrow + 2Z\downarrow],$$

$$u_6(\mathbf{r}) = |\Gamma_8, -3/2\rangle = -(1/\sqrt{2})(X - iY)\downarrow,$$

$$u_7(\mathbf{r}) = |\Gamma_7, +1/2\rangle = (1/\sqrt{3})[(X + iY)\downarrow + Z\uparrow],$$

$$u_8(\mathbf{r}) = |\Gamma_7, -1/2\rangle = (1/\sqrt{3})[(X - iY)\uparrow - Z\downarrow], \quad (s1)$$

the Kane Hamiltonian for the envelope function takes the form

$$\mathbf{H} = \begin{pmatrix} T & 0 & -\frac{1}{\sqrt{2}}Pk_+ & \sqrt{\frac{2}{3}}Pk_z & \frac{1}{\sqrt{6}}Pk_- & 0 & -\frac{1}{\sqrt{3}}Pk_z & -\frac{1}{\sqrt{3}}Pk_- \\ 0 & T & 0 & -\frac{1}{\sqrt{6}}Pk_+ & \sqrt{\frac{2}{3}}Pk_z & \frac{1}{\sqrt{2}}Pk_- & -\frac{1}{\sqrt{3}}Pk_+ & \frac{1}{\sqrt{3}}Pk_z \\ -\frac{1}{\sqrt{2}}k_-P & 0 & U+V & -\bar{S}_- & R & 0 & \frac{1}{\sqrt{2}}\bar{S}_- & -\sqrt{2}R \\ \sqrt{\frac{2}{3}}k_zP & -\frac{1}{\sqrt{6}}k_-P & -\bar{S}_-^+ & U-V & C & R & \sqrt{2}V & -\sqrt{\frac{3}{2}}\tilde{S}_- \\ \frac{1}{\sqrt{6}}k_+P & \sqrt{\frac{2}{3}}k_zP & R^+ & C^+ & U-V & \bar{S}_+^+ & -\sqrt{\frac{3}{2}}\tilde{S}_+ & -\sqrt{2}V \\ 0 & \frac{1}{\sqrt{2}}k_+P & 0 & R^+ & \bar{S}_+ & U+V & \sqrt{2}R^+ & \frac{1}{\sqrt{2}}\bar{S}_+ \\ -\frac{1}{\sqrt{3}}k_zP & -\frac{1}{\sqrt{3}}k_-P & \frac{1}{\sqrt{2}}\bar{S}_-^+ & \sqrt{2}V & -\sqrt{\frac{3}{2}}\tilde{S}_+^+ & \sqrt{2}R & U-\Delta & C \\ -\frac{1}{\sqrt{3}}k_+P & \frac{1}{\sqrt{3}}k_zP & -\sqrt{2}R^+ & -\sqrt{\frac{3}{2}}\tilde{S}_-^+ & -\sqrt{2}V & \frac{1}{\sqrt{2}}\bar{S}_+^+ & C^+ & U-\Delta \end{pmatrix}$$

where

$$T = E_c + \frac{\hbar^2}{2m_0}\left[(2F+1)(k_x^2 + k_y^2) + k_z(2F+1)k_z\right] + a_c(2\varepsilon_{xx} + \varepsilon_{zz}),$$

$$U = E_v - \frac{\hbar^2}{2m_0}\left[\gamma_1(k_x^2 + k_y^2) + k_z\gamma_1 k_z\right] + a_v(2\varepsilon_{xx} + \varepsilon_{zz}),$$

$$V = -\frac{\hbar^2}{2m_0}\left[\gamma_2(k_x^2 + k_y^2) - 2k_z\gamma_1 k_z\right] + b(\varepsilon_{xx} - \varepsilon_{zz}),$$

$$R = -\frac{\hbar^2}{2m_0}\frac{\sqrt{3}}{2}\left[(\gamma_3 - \gamma_2)k_+^2 - (\gamma_3 + \gamma_2)k_-^2\right],$$

$$\overline{S}_\pm = -\frac{\hbar^2}{2m_0}\sqrt{3}\left[k_\pm\{\gamma_3, k_z\} + k_\pm\{\kappa, k_z\}\right],$$

$$\widetilde{S}_\pm = -\frac{\hbar^2}{2m_0}\sqrt{3}\left[k_\pm\{\gamma_3, k_z\} - \frac{1}{3}k_\pm\{\kappa, k_z\}\right],$$

$$C = \frac{\hbar^2}{m_0}k_-[\kappa, k_z], \qquad k_\pm = k_x \pm ik_y, \qquad k_z = -i\partial_z. \qquad \text{(s2)}$$

Here, $[A,B] = AB - BA$ is the commutator, $\{A,B\} = AB + BA$ is the anticommutator for the operators $A$ and $B$; $P$ is the Kane momentum matrix element; $E_c(z)$ and $E_v(z)$ are the conduction and valence band edges, respectively; $\Delta(z)$ is the spin orbit energy; $a_c$ and $a_v$ are the hydrostatic and $b$ is the uniaxial deformation potentials; $\gamma_1$, $\gamma_2$, $\gamma_3$, $\kappa$ and $F$ describe the interaction with the remote bands, not considered in the Hamiltonian. The terms proportional to non-zero components of the strain tensor $\varepsilon_{xx} = \varepsilon_{yy}$ and $\varepsilon_{zz}$ result from lattice-mismatch strain. From the condition of zero external stress along the (001) direction we get the relation between $\varepsilon_{xx}$ and $\varepsilon_{zz}$:

$$\varepsilon_{xx} = \frac{a_{CdTe} - a_L}{a_L}, \qquad \varepsilon_{zz} = -\frac{2C_{12}}{C_{11}}\varepsilon_{xx}, \qquad \text{(s3)}$$

where $C_{ij}$ are the elastic constants in each layer, $a_L$ and $a_{CdTe}$ are the lattice constants of the given layer and CdTe buffer, respectively. In the Kane Hamiltonian, we omit effects of bulk inversion asymmetry introduced by the zinc blende structure of bulk HgTe and CdTe (*S1*) and contribution of spin-orbit interaction into strain-dependent part of the Hamiltonian (*S2*).

Assuming translation invariance in the *xy* plane, envelope function $F_i(\mathbf{r})$ for $u_i(\mathbf{r})$ Bloch amplitude can be represented as

$$F_i(\mathbf{r}) = \exp(ik_x x + ik_y y)f_i(z), \qquad \text{(s4)}$$

where $k_x$ and $k_y$ are the wave vector components in the QW plane. As a result, Schrödinger equation with the Kane Hamiltonian and external electric field potential $V(z)$ is reduced to the following system of differential equations:

$$\sum_{j=1}^{8}\left(\mathbf{H}_{ij}+V(z)\delta_{ij}\right)f_j(z)=\widetilde{E}_{n_z}(k_x,k_y)f_i(z), \tag{s5}$$

where $n_z$ is the electronic subband index. To solve this system, the functions $f_i(z)$ are expanded in terms of the complete basis set $\{\eta_\mu\}$ of plane waves:

$$f_i(z)=\frac{1}{\sqrt{L_z}}\sum_{\mu=-N}^{N}C_i^{(\mu)}\exp(ik_\mu z), \tag{s6}$$

where $k_\mu = 2\pi\mu/L_z$ and $L_z$ is the total width of the DQW structure in $z$ direction (in this work, $L_z = L_{CdHgTe}^{(Top)} + 2d + t + L_{CdHgTe}^{(Back)}$, and $L_{CdHgTe}^{(Back)} = L_{CdHgTe}^{(Top)} = 30$ nm). In our calculations, $N$ defines the accuracy of the solution of the eigenvalue problem, $N = 100$ is good to get convergent results with precision higher than 0.5 %.

The expansion in Eq. s6 leads to a matrix representation of the eigenvalue problem, where the eigenvectors with components $C_i^{(\mu)}$ and the corresponding eigenvalues are obtained by diagonalization of matrix $\langle\eta_\mu|\mathbf{H}_{ij}+V(z)\delta_{ij}|\eta_{\mu'}\rangle$. By using the plane-wave basis, the matrix elements $\langle\eta_\mu|K(z)|\eta_{\mu'}\rangle$, $\langle\eta_\mu|K(z)\partial_z|\eta_{\mu'}\rangle$, $\langle\eta_\mu|\partial_z K(z)|\eta_{\mu'}\rangle$ and $\langle\eta_\mu|\partial_z K(z)\partial_z|\eta_{\mu'}\rangle$ can be calculated analytically, where $K(z)$ is an arbitrary polynomial for each of the QW layers. This also allows one to calculate exactly the matrix element for external electric field potential $\langle\eta_\mu|V(z)|\eta_{\mu'}\rangle$. The term $V(z)$ is determined by,

$$\frac{d}{dz}\varepsilon(z)\frac{dV(z)}{dz}=0, \tag{s7}$$

and the boundary condition

$$\left.\frac{dV(z)}{dz}\right|_{z=-\infty}=E\cdot z, \tag{s8}$$

where $E$ is electric field strength, $\varepsilon(z)$ is the static dielectric constant, and $z = 0$ corresponds to the right CdHgTe/HgTe interface in Fig. 1.

With the basis expansion method, through the eigenvectors $\mathbf{C}$ in (s6), we can easily classify the levels. For electronic subband $n_z$, we define the relative contribution to this level from the basis states in the set $I$:

$$d_I(k_x,k_y)=\sum_{\mu=-N}^{N}\sum_{i\in I}\left|C_i^{(\mu)}(\widetilde{E}_{n_z},k_x,k_y)\right|^2, \tag{s9}$$

where $d_I(k_x,k_y)$ is so normalized that if we include all the states in the set $I$, then $d_I(k_x,k_y) = 1$. In the present work, we will calculate $d_e$ for the contribution from the $|\Gamma_6, \pm1/2\rangle$ states, $d_{lh}$ for the contribution from the $|\Gamma_8, \pm1/2\rangle$ states, $d_{so}$ for the contribution from the $|\Gamma_7, \pm1/2\rangle$ states and $d_{hh}$ for the contribution from the $|\Gamma_8, \pm3/2\rangle$ states. For example, to calculate $d_{hh}$ from (s9), we let $I$ contain $i = 3, 6$. It is clear that $d_e+d_{lh}+d_{so}+d_{hh} = 1$ at any values of $\mathbf{k}$. We classify electronic subbands in double HgTe QW as electron-like or hole-like levels by comparing the value of $d_e+d_{lh}+d_{so}$ with $d_{hh}$. The given subband is the hole-like level if $d_{hh} > d_e+d_{lh}+d_{so}$ at $\mathbf{k} = 0$.

Otherwise, the subbands are classified as electron-like, light-hole-like or spin-off-like levels, according to the dominant component in the sum $d_e+d_{lh}+d_{so}$.

To calculate the energy levels in perpendicular magnetic field $\mathbf{B} = (0, 0, B)$ we use a Peierls substitution

$$k_x = -i\frac{\partial}{\partial x} + \frac{e}{\hbar c}A_x,$$

$$k_y = -i\frac{\partial}{\partial y} + \frac{e}{\hbar c}A_y, \qquad (s10)$$

and introduce ladder operators $b^+$ and $b$:

$$b^+ = \frac{a_B}{\sqrt{2}}k_+, \qquad b = \frac{a_B}{\sqrt{2}}k_-. \qquad (s11)$$

where $a_B$ is the magnetic length ($a_B^2 = c\hbar/eB$), $e > 0$ is the elementary charge and $A$ is the magnetic vector potential in Landau gauge $A = (0, Bx, 0)$.

Additionally, the Zeeman term $H_Z$ has to be included in the Hamiltonian. According to ($S3$), $H_Z$ has the form:

$$H_Z = \mu_B B \begin{pmatrix} 1 & 0 & 0 & 0 & 0 & 0 & 0 & 0 \\ 0 & -1 & 0 & 0 & 0 & 0 & 0 & 0 \\ 0 & 0 & -3\kappa & 0 & 0 & 0 & 0 & 0 \\ 0 & 0 & 0 & -\kappa & 0 & 0 & -\sqrt{2}\kappa & 0 \\ 0 & 0 & 0 & 0 & \kappa & 0 & 0 & -\sqrt{2}\kappa \\ 0 & 0 & 0 & 0 & 0 & 3\kappa & 0 & 0 \\ 0 & 0 & 0 & -\sqrt{2}\kappa & 0 & 0 & -2\kappa & 0 \\ 0 & 0 & 0 & 0 & -\sqrt{2}\kappa & 0 & 0 & 2\kappa \end{pmatrix}, \qquad (s12)$$

where $\mu_B$ is the Bohr magneton.

We use so-called axial approximation to calculate LLs in HgTe DQW. Within this approximation we keep the in-plane rotation symmetry by omitting the warping term, proportional to $(\gamma_3 - \gamma_2)$, in $R$ (see Eq. s2). As a result, the electron wave function is written as

$$\Psi_{n,n_z,\tilde{k}}(x,y,z) = \begin{pmatrix} c_1(z,n,n_z) \cdot |n-2,\tilde{k}\rangle \\ c_2(z,n,n_z) \cdot |n-1,\tilde{k}\rangle \\ c_3(z,n,n_z) \cdot |n-3,\tilde{k}\rangle \\ c_4(z,n,n_z) \cdot |n-2,\tilde{k}\rangle \\ c_5(z,n,n_z) \cdot |n-1,\tilde{k}\rangle \\ c_6(z,n,n_z) \cdot |n,\tilde{k}\rangle \\ c_7(z,n,n_z) \cdot |n-2,\tilde{k}\rangle \\ c_8(z,n,n_z) \cdot |n-1,\tilde{k}\rangle \end{pmatrix},$$

$$|n,\widetilde{k}\rangle = \begin{cases} 0, & n<0 \\ \dfrac{\exp(i\widetilde{k}y)}{\sqrt{2^n n! \sqrt{\pi} a_B L_y}} \exp\left(-\dfrac{\widetilde{x}^2}{2a_B^2}\right) H_n\left(\dfrac{\widetilde{x}}{a_B}\right), & n \geq 0 \end{cases},$$

$$\widetilde{x} = x - \widetilde{k}a_B^2, \tag{s13}$$

where $L_y$ is the sample size along the $y$ axis, $H_n$ are the Hermitian polynomials with number $n$ ($n$ is also the Landau level index and the eigenvalue of the operator $b^+b$), $\widetilde{k}$ is the wave vector projection onto the $y$ axis.

For $n = 0$, there is one-component wave function of so-called zero-mode LLs, which are not mixed with other LLs at $n > 0$ and are formed by heavy-hole states $|\Gamma_8, -3/2\rangle$ only:

$$\Psi_{0,n_Z,\widetilde{k}}(x,y,z) = \begin{pmatrix} 0 \\ 0 \\ 0 \\ 0 \\ 0 \\ c_6(z,0,n_Z) \cdot |0,\widetilde{k}\rangle \\ 0 \\ 0 \end{pmatrix}. \tag{s14}$$

The functions $c_6(z,0,n_z)$ are satisfied to the following equations:

$$(\widetilde{H}_{n=0} + V(z))c_6(z,0,n_Z) = \widetilde{E}c_6(z,0,n_Z),$$

$$\widetilde{H}_{n=0} = E_v(z) + a_v(2\varepsilon_{xx} + \varepsilon_{zz}) + b(\varepsilon_{xx} - \varepsilon_{zz}) + \frac{\hbar^2}{2m_0}k_z\gamma_1 k_z - \frac{\hbar e}{m_0 c}\frac{\gamma_1 + \gamma_2}{2}B + 3\kappa\mu_B B. \tag{s15}$$

It is clear that the first four terms in $\widetilde{H}_{n=0}$ also define energy of hole-like subbands (H1, H2, etc.) at $\boldsymbol{k} = 0$ (cf. Eq. s2). Therefore, if the middle barrier thickness $t$ is large enough for merging of hole-like subband pairs at $\boldsymbol{k} = 0$ (see Fig. 2A, B) and $V(z) = 0$, the zero-mode LLs become double-degenerated as compared with other LLs. Thus, the unconventional quantum Hall effect measured in natural BG, should be observed in *all phases* of double HgTe QWs with $t$ values, at which H1 and H2 subbands coincide. Non-zero external electric field $V(z)$ breaks coincidence between H1 and H2 subbands at $\boldsymbol{k} = 0$ and, therefore, removes the degeneracy of zero-mode LLs, giving equidistant plateaus in the Hall conductivity.

To solve the Schrödinger equation in magnetic field, we also expand functions $c_i(z,n,n_z)$, $i = 1,\ldots 8$, by a series of plane waves, as it is done in the absence of magnetic field. In Table S1 we list the band parameters used in our calculation. The material parameters are taken from (*17*) except $a_c$, $a_v$, $b$, $d$ and the estimates for the elastic modules $C_{ij}$ which are taken from (*S4*). For the band gap $E_g = E_c - E_v$ in $Cd_xHg_{1-x}Te$ alloy, we used parabolic dependence on $x$ between the values in CdTe and HgTe (*S5*), while static dielectric constant is assumed to be $\varepsilon = 20.5 -$

15.6·$x$ + 5.7·$x^2$ (*S6*). Other band parameters are considered as piecewise functions along the growth direction and to vary linearly with $x$ in the alloy.

**Table S1.** Band parameters for HgTe and CdTe used in calculations.

| Parameters | CdTe | HgTe | Parameters | CdTe | HgTe |
|---|---|---|---|---|---|
| $E_g$, eV | 1.606 | -0.303 | $E_P$, eV | 18.8 | 18.8 |
| $E_v$, eV | -0.57 | 0 | $a$, Å | 6.48 | 6.46 |
| $\Delta$, eV | 0.91 | 1.08 | $a_c$, eV | -2.925 | -2.380 |
| $F$ | 0 | -0.09 | $a_v$, eV | 0 | 1.31 |
| $\gamma_1$ | 1.47 | 4.1 | $b$, eV | -1.2 | -1.5 |
| $\gamma_2$ | -0.28 | 0.5 | $C_{11}$, $10^{11}$ din/cm$^2$ | 5.62 | 5.92 |
| $\gamma_3$ | 0.03 | 1.3 | $C_{12}$, $10^{11}$ din/cm$^2$ | 3.94 | 4.14 |
| $\kappa$ | -1.31 | -0.4 | - | - | - |

## Effective 2D Hamiltonian for E1, E2, H1 and H2 levels

We now derive an 8×8 effective 2D Hamiltonian used for qualitative description of quantum phase transitions and corresponding picture of the edge states in double HgTe QWs. Such an approach is valid as long as 2$M_1$ and 2$M_2$ are smaller than the energy separation from considered *E*1, *E*2, *H*1, *H*2 subbands to other excited subbands. Our starting point is the eight-band Kane Hamiltonian in the absence of small terms resulted from bulk inversion asymmetry and strain-dependent part of spin-orbit interaction. Here, we focus on the case, in which structure inversion symmetry holds. Thus, the corresponding effective 2D model describes the double QWs with HgTe layers of equal thicknesses in the absence of electric field. For simplicity, we also keep the in-plane rotation symmetry by adopting the axial approximation mentioned above.

Following the procedure described in (*S7*), we split the Hamiltonian in Eq. (s2) into two parts **H** = **H**$_0$($k_z$) + **H**$_1$($k_z, k_x, k_y$), where **H**$_0$($k_z$) is the Kane Hamiltonian at $k_{x,y}$=0. First, we numerically diagonalize the Hamiltonian **H**$_0$($k_z$), to obtain the energies and envelope functions $f_i(z)$ in Eq. (s2), as well as to as to classify of electronic levels as electron-like *En*, hole-heavy-like *Hn*, light-hole-like *LHn* or spin-off-like *SOn* levels (*n* = 1, 2, …). It is clear from the form of **H**$_0$($k_z$) that the hole-heavy-like levels at $k_{x,y}$=0 are decoupled from the *En*, *LHn* and *SOn* subbands. Therefore, the eigenfunctions of **H**$_0$($k_z$) are expanded in the basis of Bloch amplitudes as follows:

$$|En,+\rangle = f_1^{(En,+)}(z)|\Gamma_6,+1/2\rangle + f_4^{(En,+)}(z)|\Gamma_8,+1/2\rangle + f_7^{(En,+)}(z)|\Gamma_7,+1/2\rangle,$$

$$|En,-\rangle = f_2^{(En,-)}(z)|\Gamma_6,-1/2\rangle + f_5^{(En,-)}(z)|\Gamma_8,-1/2\rangle + f_8^{(En,-)}(z)|\Gamma_7,-1/2\rangle,$$

$$|Hn,+\rangle = f_3^{(Hn,+)}(z)|\Gamma_8,+3/2\rangle,$$

$$|Hn,-\rangle = f_6^{(Hn,-)}(z)|\Gamma_8,-3/2\rangle$$

$$|LHn,+\rangle = f_1^{(LHn,+)}(z)|\Gamma_6,+1/2\rangle + f_4^{(LHn,+)}(z)|\Gamma_8,+1/2\rangle + f_7^{(LHn,+)}(z)|\Gamma_7,+1/2\rangle,$$

$$|LHn,-\rangle = f_2^{(LHn,-)}(z)|\Gamma_6,-1/2\rangle + f_5^{(LHn,-)}(z)|\Gamma_8,-1/2\rangle + f_8^{(LHn,-)}(z)|\Gamma_7,-1/2\rangle,$$

$$|SOn,+\rangle = f_1^{(SOn,+)}(z)|\Gamma_6,+1/2\rangle + f_4^{(SOn,+)}(z)|\Gamma_8,+1/2\rangle + f_7^{(SOn,+)}(z)|\Gamma_7,+1/2\rangle,$$

$$|SOn,-\rangle = f_2^{(SOn,-)}(z)|\Gamma_6,-1/2\rangle + f_5^{(SOn,-)}(z)|\Gamma_8,-1/2\rangle + f_8^{(SOn,-)}(z)|\Gamma_7,-1/2\rangle. \tag{s16}$$

Since we hold structure inversion symmetry, the inversion operation **P**, defining the parity of each subband, commutes with the Hamiltonian $\mathbf{H}_0(k_z)$. The parity of the subbands is determined by both the envelope functions $f_i(z)$ and the Bloch amplitudes in the $\Gamma$ point. The parities of the envelope functions are obtained through numerical calculations, and are given in Table S2. The parities of the Bloch amplitudes are given by $\mathbf{P}|\Gamma_6, \pm 1/2\rangle = -|\Gamma_6, \pm 1/2\rangle$, $\mathbf{P}|\Gamma_7, \pm 1/2\rangle = |\Gamma_7, \pm 1/2\rangle$, $\mathbf{P}|\Gamma_8, \pm 1/2\rangle = |\Gamma_8, \pm 1/2\rangle$ and, $\mathbf{P}|\Gamma_8, \pm 3/2\rangle = |\Gamma_8, \pm 3/2\rangle$. Thus, the parities of the subbands are $\mathbf{P}|E1,\pm\rangle = -|E1,\pm\rangle$, $\mathbf{P}|E2,\pm\rangle = |E2,\pm\rangle$, $\mathbf{P}|H1,\pm\rangle = |H1,\pm\rangle$, $\mathbf{P}|H2,\pm\rangle = -|H2,\pm\rangle$, $\mathbf{P}|LH1,\pm\rangle = |LH1,\pm\rangle$, $\mathbf{P}|LH2,\pm\rangle = -|LH2,\pm\rangle$, and etc.

**Table S2.** Parities of the envelope function components.

| Subband | E2,+ | E2,− | E1,+ | E1,− | LH1,+ | LH1,− |
|---|---|---|---|---|---|---|
| Even | $f_4^{(E2,+)}(z)$, $f_7^{(E2,+)}(z)$ | $f_5^{(E2,-)}(z)$, $f_8^{(E2,-)}(z)$ | $f_1^{(E1,+)}(z)$ | $f_2^{(E1,-)}(z)$ | $f_4^{(LH1,+)}(z)$, $f_7^{(LH1,+)}(z)$ | $f_5^{(LH1,-)}(z)$, $f_8^{(LH1,-)}(z)$ |
| Odd | $f_1^{(E2,+)}(z)$ | $f_2^{(E2,-)}(z)$ | $f_4^{(E1,+)}(z)$, $f_7^{(E1,+)}(z)$ | $f_5^{(E1,-)}(z)$, $f_8^{(E1,-)}(z)$ | $f_1^{(LH1,+)}(z)$ | $f_2^{(LH1,-)}(z)$ |

| Subband | H(2n),+ | H(2n),− | H(2n-1),+ | H(2n-1),− |
|---|---|---|---|---|
| Even | − | − | $f_3^{(H(2n-1),+)}(z)$ | $f_6^{(H(2n-1),-)}(z)$ |
| Odd | $f_3^{(H(2n),+)}(z)$ | $f_6^{(H(2n),-)}(z)$ | − | − |

In accordance with (S7), we group the eigenstates of Eq. s16 into two classes. The first class one, marked as class A, includes the basis states of our final effective 2D Hamiltonian $\{|E1,\pm\rangle, |H1,\pm\rangle, |H2,\pm\rangle, |E2,\pm\rangle\}$. In the second class, denoted as class B, we consider the following basis states $\{|H3,\pm\rangle, |H4,\pm\rangle, |LH1,\pm\rangle, |LH2,\pm\rangle, |H5,\pm\rangle, |H6,\pm\rangle\}$. All the other subbands of the QW are neglected since they are well separated in energy. The states in both classes are not coupled, since they are eigenstates of Hamiltonian $\mathbf{H}_0(k_z)$. However, the presence of $\mathbf{H}_1(k_z, k_x, k_y)$ introduces the mixing between the states from classes A and B. To derive an effective 2D Hamiltonian $H_{eff}(k_x, k_y)$ for double HgTe, we treat $\mathbf{H}_1(k_z, k_x, k_y)$ as a small perturbation and perform a unitary transformation (S3) to eliminate the coupling between the states from different classes by applying the second-order perturbation formula

$$H_{eff}(k_x,k_y)_{m,m'} = E_m \delta_{m,m'} + H'_{m,m'} + \frac{1}{2}\sum_l H'_{m,l} H'_{l,m'}\left(\frac{1}{E_m - E_l} + \frac{1}{E_{m'} - E_l}\right), \quad (s17)$$

where

$$E_m = \int_{-\infty}^{+\infty} dz \sum_{\alpha,\beta=1}^{8} f_\alpha^{(m)*}(z)\left(\mathbf{H}_0(k_z)\right)f_\beta^{(m)}(z),$$

$$H'_{m,m'} = \int_{-\infty}^{+\infty} dz \sum_{\alpha,\beta=1}^{8} f_\alpha^{(m)*}(z)\left(\mathbf{H}_1(k_z,k_x,k_y)\right)f_\beta^{(m')}(z). \quad (s18)$$

The summation indices $m$, $m'$ correspond to the states in class A, while index $l$ is for the states in class B. The Greek indices label envelope function component of the Kane Hamiltonian. We note that accounting for the parity of the envelope functions $f_\alpha^{(m)}(z)$, given in Table S2, greatly simplifies calculation of $H'_{m,m'}$ in the perturbation procedure above.

Ordering the basis states as $\{|E1,+\rangle, |H1,+\rangle, |H2,-\rangle, |E2,-\rangle, |E2,+\rangle, |H2,+\rangle, |H1,-\rangle, |E1,-\rangle\}$, after calculating the matrix-elements (s17), we are left with an effective Hamiltonian parameterized in the following way

$$H_{eff}(k_x,k_y) = \begin{pmatrix} H(k_x,k_y) & 0 \\ 0 & \Theta H(k_x,k_y)\Theta^{-1} \end{pmatrix}, \quad (s19)$$

where $\Theta$ is a "time reversal" operator, given by

$$\Theta = \begin{pmatrix} 0 & i\sigma_y \\ i\sigma_y & 0 \end{pmatrix} K, \quad (s20)$$

with $K$ stands for complex conjugation and $\sigma_y$ is one of the Pauli spin matrices. In (s19) $H(k_x, k_y)$ is written as

$$H(k_x,k_y) = \begin{pmatrix} \varepsilon_{E1}(k) & -A_1 k_+ & R_1 k_-^2 & S_0 k_- \\ -A_1 k_- & \varepsilon_{H1}(k) & 0 & R_2 k_-^2 \\ R_1 k_+^2 & 0 & \varepsilon_{H2}(k) & A_2 k_+ \\ S_0 k_+ & R_2 k_+^2 & A_2 k_- & \varepsilon_{E2}(k) \end{pmatrix}, \quad (s21)$$

$$k_\pm = k_x + ik_y,$$

$$\varepsilon_{E1}(k) = C + \frac{\Delta_{H1H2}}{2} + 2M_1 + B_{E1}(k_x^2 + k_y^2),$$

$$\varepsilon_{H1}(k) = C + \frac{\Delta_{H1H2}}{2} + B_{H1}(k_x^2 + k_y^2),$$

$$\varepsilon_{H2}(k) = C - \frac{\Delta_{H1H2}}{2} + B_{H2}(k_x^2 + k_y^2),$$

$$\varepsilon_{E2}(k) = C - \frac{\Delta_{H1H2}}{2} + 2M_2 + B_{E2}(k_x^2 + k_y^2).$$

Band parameters $C$, $M_1$, $M_2$, $A_1$, $A_2$, $B_{E1}$, $B_{H1}$, $B_{H2}$, $B_{E2}$, $\Delta_{H1H2}$, $R_1$, $R_2$, $S_0$ depend on $d$, $t$ and material of the QW and barrier layers and are calculated numerically by using envelope function component (integration over $z$ axis). Their values are listed in Table S3. We note that if $\Delta_{H1H2} = 0$ the straightforward calculations results in $B_{H1} = B_{H2}$. Comparison of the bulk electronic subbands, calculated within the eight-band Kane model and Hamiltonian $H_{\mathit{eff}}(k_x, k_y)$, is given in Fig. S3. One can see good agreement between results from both models at small quasimomentum values.

**Table S3.** Structure parameters involved in effective Hamiltonian $H(k_x, k_y)$.

| Panel in Fig. 4 | $d$, nm | $t$, nm | $C$, meV | $\Delta_{H1H2}$, meV | $M_1$, meV | $M_2$, meV | $A_1$, meV·nm | $A_2$, meV·nm |
|---|---|---|---|---|---|---|---|---|
| A | 6.0 | 6.65 | −31.1 | 0 | 0 | 13.8 | 387 | −373 |
| B | 6.5 | 3.0 | −25.1 | 0 | −11.3 | 8.8 | 378 | 358 |
| C | 7.5 | 3.28 | −24.2 | 0 | −16.1 | 0 | 358 | −340 |
| D | 7.5 | 4.0 | −21.4 | 0 | −16.3 | −2.25 | 360 | −348 |

| Panel in Fig. 4 | $R_1$, meV·nm$^2$ | $R_2$, meV·nm$^2$ | $S_0$, meV·nm | $B_{E1}$, meV·nm$^2$ | $B_{H1}$, meV·nm$^2$ | $B_{H2}$, meV·nm$^2$ | $B_{E2}$, meV·nm$^2$ |
|---|---|---|---|---|---|---|---|
| A | 238 | 101 | 0.5 | 1258 | −145 | −145 | 893 |
| B | −160 | 152 | −6.5 | 1758 | −115 | −115 | 894 |
| C | 478 | 154 | −7.3 | 2302 | −109 | −109 | 1096 |
| D | −329 | −107 | 2.3 | 2138 | −152 | −152 | 1253 |

The form (s20) for a "time reversal" operator $\Theta$ mimics the one for spin $J = 3/2$. If one rewrites the quasimomentum in the polar coordinate system as $k_x = k\cdot\cos(\theta)$, $k_y = k\cdot\sin(\theta)$, it is easy to make sure that the following equation holds

$$H(k,0) = \exp(i \cdot J_z \theta) \cdot H(k\cos(\theta), k\sin(\theta)) \cdot \exp(-i \cdot J_z \theta), \quad (s22)$$

where $J_z$ is one of the spin matrices:

$$J_x = \frac{1}{2}\begin{pmatrix} 0 & \sqrt{3} & 0 & 2 \\ \sqrt{3} & 0 & 0 & 0 \\ 0 & 0 & 0 & \sqrt{3} \\ 2 & 0 & \sqrt{3} & 0 \end{pmatrix}, \quad J_y = \frac{i}{2}\begin{pmatrix} 0 & \sqrt{3} & 0 & -2 \\ -\sqrt{3} & 0 & 0 & 0 \\ 0 & 0 & 0 & \sqrt{3} \\ 2 & 0 & -\sqrt{3} & 0 \end{pmatrix}, \quad J_z = \frac{1}{2}\begin{pmatrix} 1 & 0 & 0 & 0 \\ 0 & 3 & 0 & 0 \\ 0 & 0 & -3 & 0 \\ 0 & 0 & 0 & -1 \end{pmatrix}. \quad (s23)$$

Here the matrices are written for the following order of $J_z = \{1/2, 3/2, -3/2, -1/2\}$.

Equation (s22) means that the wave function $|S_k^\theta\rangle$ for the upper block in $H_{eff}(k\cdot\cos(\theta), k\cdot\sin(\theta))$ can be obtained from the initial state along the $x$-axis by the rotation operation $|S_k^\theta\rangle = \exp(-i \cdot J_z \theta)|S_k^0\rangle$. The latter clearly resembles that of a four-component spinor describing the spin $J = 3/2$, but arising from the symmetry of E1, H1, H2, E2 subbands. We note that the above rotation operation also implies that the orientation of the pseudo spin $\mathbf{J} = (J_x, J_y, J_z)$ is tied to the quasimomentum $\mathbf{k}_\| = (k_x, k_y, 0)$. So far our analysis is focused on the upper block of $H_{eff}(k_x, k_y)$. Since $\Theta$ commutes with $\exp(-i\cdot J_z\theta)$, the rotation operation can be applied for spinor of $\Theta H_{eff}(k_x, k_y)\Theta^{-1}$ as well.

Using (s23) and the anticommutator $\{J_\alpha, J_\beta\} = J_\alpha J_\beta + J_\beta J_\alpha$, we can alternatively write for $H(k_x, k_y)$

$$H(k_x, k_y) = \begin{pmatrix} \tilde{\varepsilon}_{E1}(k) & 0 & 0 & 0 \\ 0 & \tilde{\varepsilon}_{H1}(k) & 0 & 0 \\ 0 & 0 & \tilde{\varepsilon}_{H2}(k) & 0 \\ 0 & 0 & 0 & \tilde{\varepsilon}_{E2}(k) \end{pmatrix} + H_k + H_{k^2}, \quad (s24)$$

$$H_k = \frac{1}{4}\left(\frac{A_1 - A_2}{\sqrt{3}} + 5S_0\right)(\vec{k}\cdot\vec{J}) - \frac{1}{2}\frac{A_1 + A_2}{\sqrt{3}}\left(k_x\{J_x, J_z\} + k_y\{J_y, J_z\}\right) -$$
$$- \frac{1}{2}\left(\frac{A_1 - A_2}{\sqrt{3}} + S_0\right)\left(k_x\{J_x, J_z^2\} + k_y\{J_y, J_z^2\}\right),$$

$$H_{k^2} = \frac{R_1 + R_2}{\sqrt{3}}(\vec{k}\cdot\vec{J})^2 + \frac{R_2 - R_1}{\sqrt{3}}\left((k_x^2 - k_y^2)(J_x^2 J_z - J_z J_y^2) + 2k_x k_y (J_x J_y J_z + J_z J_y J_x)\right).$$

where $\tilde{\varepsilon}_{E1}(k), \tilde{\varepsilon}_{H1}(k), \tilde{\varepsilon}_{H2}(k), \tilde{\varepsilon}_{E2}(k)$ are obtained from $\varepsilon_{E1}(k), \varepsilon_{H1}(k), \varepsilon_{H2}(k), \varepsilon_{E2}(k)$ by the following substitution:

$$B_{E1} = \tilde{B}_{E1} + \frac{7\sqrt{3}}{12}(R_1 + R_2), \qquad B_{H1} = \tilde{B}_{H1} + \frac{\sqrt{3}}{4}(R_1 + R_2),$$

$$B_{H2} = \widetilde{B}_{H2} + \frac{\sqrt{3}}{4}(R_1 + R_2), \qquad B_{E2} = \widetilde{B}_{E2} + \frac{7\sqrt{3}}{12}(R_1 + R_2).$$

Expression for $\Theta H_{eff}(k_x, k_y)\Theta^{-1}$ via the spin 3/2 matrices is obtained from (s24) by the substitution $\mathbf{J} \to -\mathbf{J}$. Therefore, the electron wave function corresponding to the down block of $H_{eff}(k_x, k_y)$ is also characterized by fermions described by $H(k_x, k_y)$ but with opposite helicity.

The Hamiltonian (s24) looks rather complicated; however, it contains the minimal number of quantities for description of BI, TI and BG phases in DQW HgTe. We note that further simplification, performed, for an example, for BG phase by projecting $H(k_x, k_y)$ onto the basic functions of $H1$ and $H2$ subbands at $\mathbf{k}_\parallel = 0$, results in the model, which actually does not describe behaviour of LLs in the vicinity of Γ point. In particular, the coincidence between zero-mode LLs, marked as $H1$ and $H2$ in Fig. 3D, is missed in this case. The latter can only be described within the Hamiltonian, directly accounting for the coupling between $E1$, $H1$, $H2$, $E2$ subbands.

## Numerical calculation of edge state dispersions

To consider the edge states on a single edge, we deal with a system on a half-plane of $y \leq 0$ and replace $k_y$ by $-i\partial_y$ in this case. If $M_1 < 0$, the DQW structure supports edge states, which exponentially decay at $y \to -\infty$. The Hamiltonian $H_{eff}(k_x, k_y)$ (s19) is block diagonal, and the eigenvalue problem of the upper and lower blocks can be solved separately. To find the energy spectrum the edge states for the upper block, we solve the Schrödinger equation:

$$H(k_x, -i\partial_y)\Psi_{edge}(x, y) = E_{edge}\Psi_{edge}(x, y) \tag{s25}$$

and put the boundary condition for the wave function to vanish at $y = 0$. Taking into account the translation invariance along the x axis, the wave function of the edge states has the form

$$\Psi_{edge}(x, y) \frac{e^{ik_x x}}{\sqrt{L_x}} \sum_{n=1}^{4} \alpha_n e^{\lambda_n y} C_n^{(edge)}, \tag{s26}$$

where $k_x$ is the wave vector along the edge, $L_x$ is the sample size along the $x$ axis, $\alpha_n$ are the coefficients, determined by the boundary conditions, $\lambda_n$ are the complex-valued reciprocal lengths, and $C_n^{(edge)}$ are the position-independent normalized four-component columns.

For a given wave vector $k_x$, relation $\lambda_n(E_{edge})$ and columns $C_n^{(edge)}$ are found from the matrix equation

$$H(k_x, -i\lambda)C^{(edge)} = E_{edge}C^{(edge)}. \tag{s27}$$

Note that $\lambda_n$ in general can be complex, corresponding to a mixture of the edge and bulk states. The energy spectrum of the edge states are found from the condition of wave function decay at $y \to -\infty$ (by implying Re $\lambda_n > 0$) and from the boundary condition at $y \to 0$:

$$\sum_{n=1}^{4} \alpha_n C_n^{(edge)} = 0. \tag{s28}$$

The energy spectrum of the edge states for the lower block in $H_{eff}(k_x, k_y)$ is found in a similar way. Fig. 4 illustrates dispersions of the bulk and edge states in the vicinity of the Γ point, calculated by using $H_{eff}(k_x, k_y)$ Hamiltonian.

## Effective model of Michetti et al.

The first attempt to investigate properties of DQW HgTe has been performed by Michetti et al. (*14,15*). For the case of well-separated HgTe layers of thickness close to $d_c$, they proposed to describe each QW with the BHZ model with an additional tunneling Hamiltonian for interlayer coupling.

According to their approach, the functions for electron-like and hole-like states {$|Ef,+\rangle$, $|Hf,+\rangle$, $|Eb,+\rangle$, $|Hb,+\rangle$, $|Ef,-\rangle$, $|Hf,-\rangle$, $|Eb,-\rangle$, $|Hb,-\rangle$}, localized in front (*f*) and (*b*) back HgTe layer, are considered as basis states at $k_\parallel = 0$. In this basis, their effective 8×8 Hamiltonian for DQW has the form:

$$H_M(k_x, k_y) = \begin{pmatrix} H_0^{(f+)} & h_T^{(+)} & 0 & 0 \\ \tilde{h}_T^{(+)} & H_0^{(b+)} & 0 & 0 \\ 0 & 0 & H_0^{(f-)} & h_T^{(-)} \\ 0 & 0 & \tilde{h}_T^{(-)} & H_0^{(b-)} \end{pmatrix}, \quad (s29)$$

where tilde means the Hermitian conjugation; $H_0^{(f+)}, H_0^{(f-)}, H_0^{(b+)}, H_0^{(b-)}$ are 2×2 BHZ Hamiltonians for the pair of $|Ef,\pm\rangle$, $|Hf,\pm\rangle$ levels and for the ones of $|Eb,\pm\rangle$, $|Hb,\pm\rangle$ respectively; $h_T$ is a tunneling Hamiltonian, written as:

$$h_T = \begin{pmatrix} \Delta_E & \alpha k_+ \\ \alpha k_- & \Delta_H \end{pmatrix}, \quad (s30)$$

with

$$\Delta_E = 2\langle Ef+|\mathbf{H}|Eb+\rangle,$$

$$\Delta_H = 2\langle Hf+|\mathbf{H}|Hb+\rangle,$$

$$\alpha k_+ = 2\langle Ef+|\mathbf{H}|Hb+\rangle. \quad (s31)$$

Here, **H** is the eight-band Kane Hamiltonian for DQW.

The essential problem of the model presented above is that the functions $|Ef,+\rangle$, $|Hf,+\rangle$, $|Eb,+\rangle$, $|Hb,+\rangle$, $|Ef,-\rangle$, $|Hf,-\rangle$, $|Eb,-\rangle$, $|Hb,-\rangle$ localized in the (*f*) and (*b*) QWs, do not form the orthogonal basis in the case of coupling between layers. It is clear that the inversion operation **P** should commute with the Hamiltonian for symmetric DQW. Therefore, the wave functions corresponding to the two consecutive levels should have different parities and should be the eigenfunctions of the operation **P**. The latter holds even if these two levels have the same energy.

The states $|Ef,\pm\rangle$ and $|Eb,\pm\rangle$, as well as $|Hf,\pm\rangle$ and $|Hb,\pm\rangle$ have the same parity, otherwise (s30) does not occur. Let us demonstrate it. It is clear that

$$|Ef,+\rangle = f_1^{(Ef,+)}(z)|\Gamma_6,+1/2\rangle + f_4^{(Ef,+)}(z)|\Gamma_8,+1/2\rangle + f_7^{(Ef,+)}(z)|\Gamma_7,+1/2\rangle,$$

$$|Eb,+\rangle = f_1^{(Eb,+)}(z)|\Gamma_6,+1/2\rangle + f_4^{(Eb,+)}(z)|\Gamma_8,+1/2\rangle + f_7^{(Eb,+)}(z)|\Gamma_7,+1/2\rangle,$$

$$|Hf,+\rangle = f_3^{(Hf,+)}(z)|\Gamma_8,+3/2\rangle,$$

$$|Hb,+\rangle = f_3^{(Hb,+)}(z)|\Gamma_8,+3/2\rangle,$$

where $f_1^{(Ef,+)}(z)$, $f_1^{(Eb,+)}(z)$, $f_3^{(Hf,+)}(z)$, $f_3^{(Hb,+)}(z)$ have even parity, while the envelope functions $f_4^{(Ef,+)}(z)$, $f_4^{(Eb,+)}(z)$, $f_7^{(Ef,+)}(z)$, $f_7^{(Eb,+)}(z)$ are of odd parity. This is also illustrated by Fig. 5 in (*14*). Straightforward calculation with accounting of (s2) results in the following expressions for the matrix elements:

$$\langle Ef+|\mathbf{H}|Eb+\rangle = \int_{-\infty}^{\infty} f_1^{(Ef,+)*}(z) T f_1^{(Eb,+)}(z) dz + \int_{-\infty}^{\infty} f_4^{(Ef,+)*}(z)(U-V) f_4^{(Eb,+)}(z) dz +$$

$$+ \int_{-\infty}^{\infty} f_7^{(Ef,+)*}(z)(U-\Delta) f_7^{(Eb,+)}(z) dz,$$

$$\langle Hf+|\mathbf{H}|Hb+\rangle = \int_{-\infty}^{\infty} f_3^{(Ef,+)*}(z)(U+V) f_3^{(Eb,+)}(z) dz,$$

$$\langle Ef+|\mathbf{H}|Hb+\rangle = -\int_{-\infty}^{\infty} f_1^{(Ef,+)*}(z) \frac{Pk_+}{\sqrt{2}} f_3^{(Hb,+)}(z) dz - \int_{-\infty}^{\infty} f_4^{(Ef,+)*}(z) \overline{S}_-^+ f_3^{(Hb,+)}(z) dz +$$

$$+ \int_{-\infty}^{\infty} f_7^{(Ef,+)*}(z) \frac{\overline{S}_-^+}{\sqrt{2}} f_3^{(Hb,+)}(z) dz. \quad (s32)$$

We note that $\int_{-\infty}^{\infty} f_1^{(Ef,+)*}(z) \frac{Pk_+}{\sqrt{2}} f_3^{(Hb,+)}(z) dz = 0$ because the integrand is an odd function of $z$. Taking into account expressions for $T$, $U$, $V$, $\Delta$ and $\overline{S}_\pm$ (s2), it is easy to see that

$$\langle Ef+|\mathbf{H}|Eb+\rangle = \frac{\Delta_E}{2} + \alpha_E(k_x^2 + k_y^2),$$

$$\langle Hf+|\mathbf{H}|Hb+\rangle = \frac{\Delta_H}{2} + \alpha_H(k_x^2 + k_y^2),$$

$$\langle Ef+|\mathbf{H}|Hb+\rangle = \frac{\alpha k_+}{2}, \quad (s33)$$

where $\Delta_E$, $\Delta_H$, $\alpha_E$, $\alpha_H$, $\alpha$ are independent on quasimomentum. Up to the second order in $k_\parallel$, expressions (s31) and (s33) coincide, proving that basic functions $|Ef,\pm\rangle$ and $|Eb,\pm\rangle$, as well as $|Hf,\pm\rangle$, and $|Hb,\pm\rangle$ have the same parity. As we have mentioned above, the latter means that the functions $|Ef,+\rangle$, $|Hf,+\rangle$, $|Eb,+\rangle$, $|Hb,+\rangle$, $|Ef,-\rangle$, $|Hf,-\rangle$, $|Eb,-\rangle$, $|Hb,-\rangle$ do not generate the orthogonal basis.

To overcome this contradiction, one should redefine the basic functions, formed by symmetrical and antisymmetrical combination of initial states $|Ef,\pm\rangle$, $|Hf,\pm\rangle$, $|Eb,\pm\rangle$ and $|Hb,\pm\rangle$:

$$|E1,\pm\rangle = \frac{|Ef\pm\rangle - |Eb\pm\rangle}{2},$$

$$|E2,\pm\rangle = \frac{|Ef\pm\rangle + |Eb\pm\rangle}{2},$$

$$|H1,\pm\rangle = \frac{|Hf\pm\rangle + |Hb\pm\rangle}{2},$$

$$|H2,\pm\rangle = \frac{|Hf\pm\rangle - |Hb\pm\rangle}{2}.$$

As it can be demonstrated, applying this basis to matrix element calculations, similar to (s32), one obtains the Hamiltonian $H_{eff}(k_x,k_y)$ given by (s21).

## Edge states in two tunnel-coupled layers of 2D TI

DQW HgTe at $d > d_c$ and large values of $t$ can be considered as two tunnel-coupled layers of 2D TI. In this case, it is educative to rewrite $H_{eff}(k_x,k_y)$ in the following way:

$$H_{eff}(k_x,k_y) = \hat{H}_0(k_x,k_y) + \hat{H}_T^{(0)}(k_x,k_y) + \hat{H}_T^{(SF)}(k_x,k_y), \quad (s34)$$

where the first term describes the states in two separated HgTe layers

$$\hat{H}_0(k_x,k_y) = \begin{pmatrix} H_0(k_x,k_y) & 0 & 0 & 0 \\ 0 & \Theta_{2D} H_0(k_x,k_y)\Theta_{2D}^{-1} & 0 & 0 \\ 0 & 0 & H_0(k_x,k_y) & 0 \\ 0 & 0 & 0 & \Theta_{2D} H_0(k_x,k_y)\Theta_{2D}^{-1} \end{pmatrix} \quad (s35)$$

with $\Theta_{2D} = \exp(-i\pi\sigma_y/2)$ and 2×2 BHZ Hamiltonian $H_0(k_x,k_y)$. The block matrices on the diagonal are written for the following order of the levels: $\{|E1,+\rangle, |H1,+\rangle\}$, $\{|H2,-\rangle, |E2,-\rangle\}$, $\{|E2,+\rangle, |H2,+\rangle\}$, $\{|H1,-\rangle, |E1,-\rangle\}$.

Two other terms in (s34) are interpreted as spin-conserved $\hat{H}_T^{(0)}(k_x,k_y)$ and spin-dependent $\hat{H}_T^{(SF)}(k_x,k_y)$ tunneling. They are written as:

$$\hat{H}_T^{(0)}(k_x,k_y) = \begin{pmatrix} -\delta H_0(k_x,k_y) & 0 & 0 & 0 \\ 0 & \Theta_{2D}\delta H_0(k_x,k_y)\Theta_{2D}^{-1} & 0 & 0 \\ 0 & 0 & \delta H_0(k_x,k_y) & 0 \\ 0 & 0 & 0 & -\Theta_{2D}\delta H_0(k_x,k_y)\Theta_{2D}^{-1} \end{pmatrix},$$

$$\hat{H}_T^{(SF)}(k_x,k_y) = \begin{pmatrix} 0 & h_T^{(SF)}(k_x,k_y) & 0 & 0 \\ \left(h_T^{(SF)}(k_x,k_y)\right)^+ & 0 & 0 & 0 \\ 0 & 0 & 0 & \left(\Theta_{2D} h_T^{(SF)}(k_x,k_y)\Theta_{2D}^{-1}\right)^+ \\ 0 & 0 & \Theta_{2D} h_T^{(SF)}(k_x,k_y)\Theta_{2D}^{-1} & 0 \end{pmatrix},$$

where "+" refers to the Hermitian conjugation. The explicit forms for $\hat{H}_0(k_x,k_y)$, $\hat{H}_T^{(0)}(k_x,k_y)$ and $\hat{H}_T^{(SF)}(k_x,k_y)$ are obtained from comparison with (s19-s21). Here, we are interesting in qualitative changes in the energy spectrum of the edge states in a clean system if one accounts $\hat{H}_T^{(0)}(k_x,k_y)$ and $\hat{H}_T^{(SF)}(k_x,k_y)$ consistently.

In the absence of tunneling, the edge states in double QW consist of two coinciding cones, each resulting from isolated HgTe QW with negative mass parameter. In the presence of spin-conserved tunneling, the cones are splitted into symmetric and anti-symmetric states, as it is schematically shown in Fig. S4A. We note that $\hat{H}_T^{(0)}(k_x,k_y)$ just renormalizes the parameters of BHZ Hamiltonians, standing on the diagonal of $\hat{H}_0(k_x,k_y)$ (s35). The coupling between different spin states from different layers, which is described by $\hat{H}_T^{(SF)}(k_x,k_y)$, leads to the gap opening (in Fig. S4B). In this case, Kramer's doublets are formed by combination of different spin states from different HgTe layers, as shown in Fig. S4C.

S1. M. H. Weiler, in *Defects, (HgCd)Se, (HgCd)Te*, R. K. Willardson and A. C. Beer, Eds. (Academic Press, New York, 1981), vol. 16, p. 119.

S2. T. B. Bahder, Eight-band **k·p** model of strained zinc-blende crystals. *Phys. Rev. B* **41**, 11992 (1990).

S3. R. Winkler, in *Spin-Orbit Coupling Effects in Two-Dimensional Electron and Hole Systems*, Springer Tracts in Modern Physics (Springer, Berlin, 2003), vol. 191.

S4. K. Takita, K. Onabe, S. Tanaka, Anomalous magnetoresistance and band crossing in uniaxially compressed HgTe. *Phys. Status Solidi (b)* **92**, 297 (1979).

S5. P. Laurenti, *et al.*, Temperature dependence of the fundamental absorption edge of mercury cadmium telluride. *J. Appl. Phys.* **67**, 6454 (1990).

S6. J. D. Patterson, W. A. Gobba, S.L. Lehoczky, Electron mobility in n-type $Hg_{1-x}Cd_x Te$ and $Hg_{1-x}Zn_x Te$ alloys. *J. Mater. Res.* **7**, 2211 (1992).

S7. D. G. Rothe, *et al.*, Fingerprint of different spin–orbit terms for spin transport in HgTe quantum wells. *New J. Phys.* **12**, 065012 (2010).

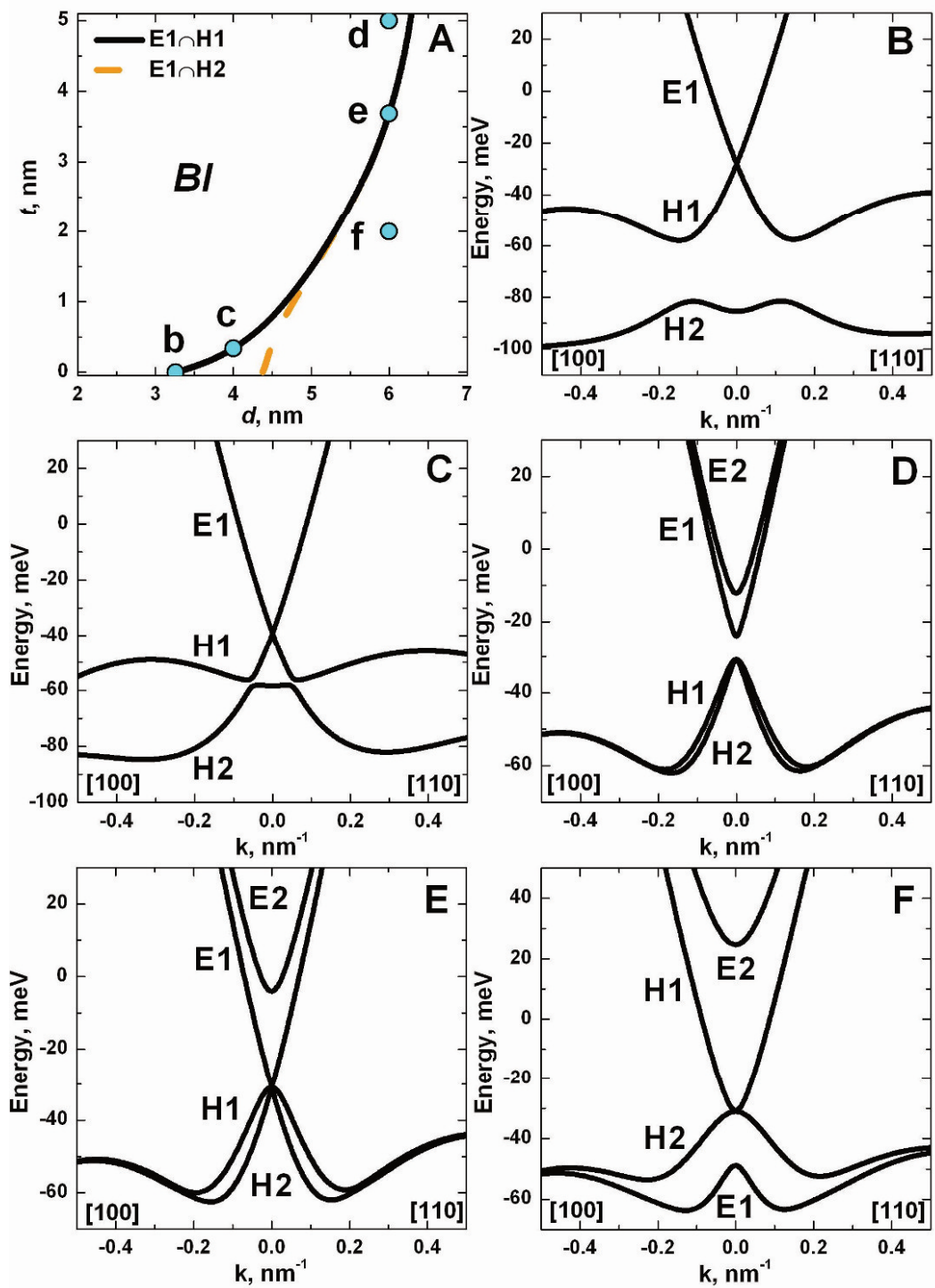

**Fig. S1**. (**A**) A part of the phase diagram showing the QW parameters, at which E1 and H1 subbands are crossed at $k = 0$ (bold black curve). The dashed orange curve corresponds to the crossing between E1 and H2 subbands in the Γ point. (**B-F**) Energy dispersions at various $d$ and $t$, which values are shown in the panel (**A**) by blue symbols and marked from b to f respectively.

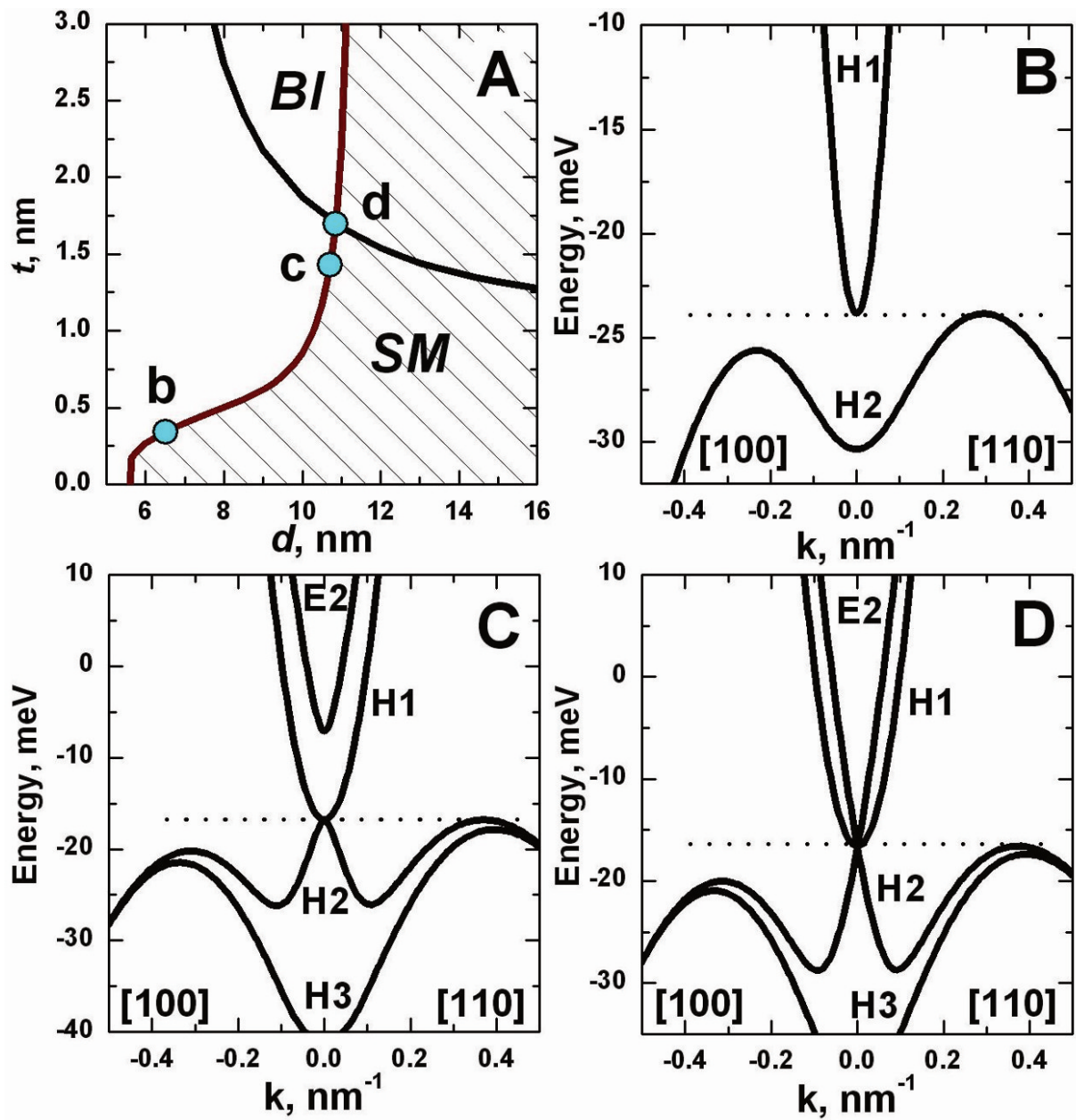

**Fig. S2**. (**A**) A part of the phase diagram, which demonstrates transition into SM phase. The QW parameters, at which a side maximum in the valence band has the same energy as a conduction band bottom, are shown by brown curve. The black curve corresponds to the crossing between $E2$ and $H2$ subbands in the Γ point. (**B-D**) Energy dispersions for the values of $d$ and $t$, shown in the panel (**A**) by blue symbols and marked by $b$, $c$ and $d$ respectively. First electron-like subband $E1$ in all the panels lies significantly below the energy scale.

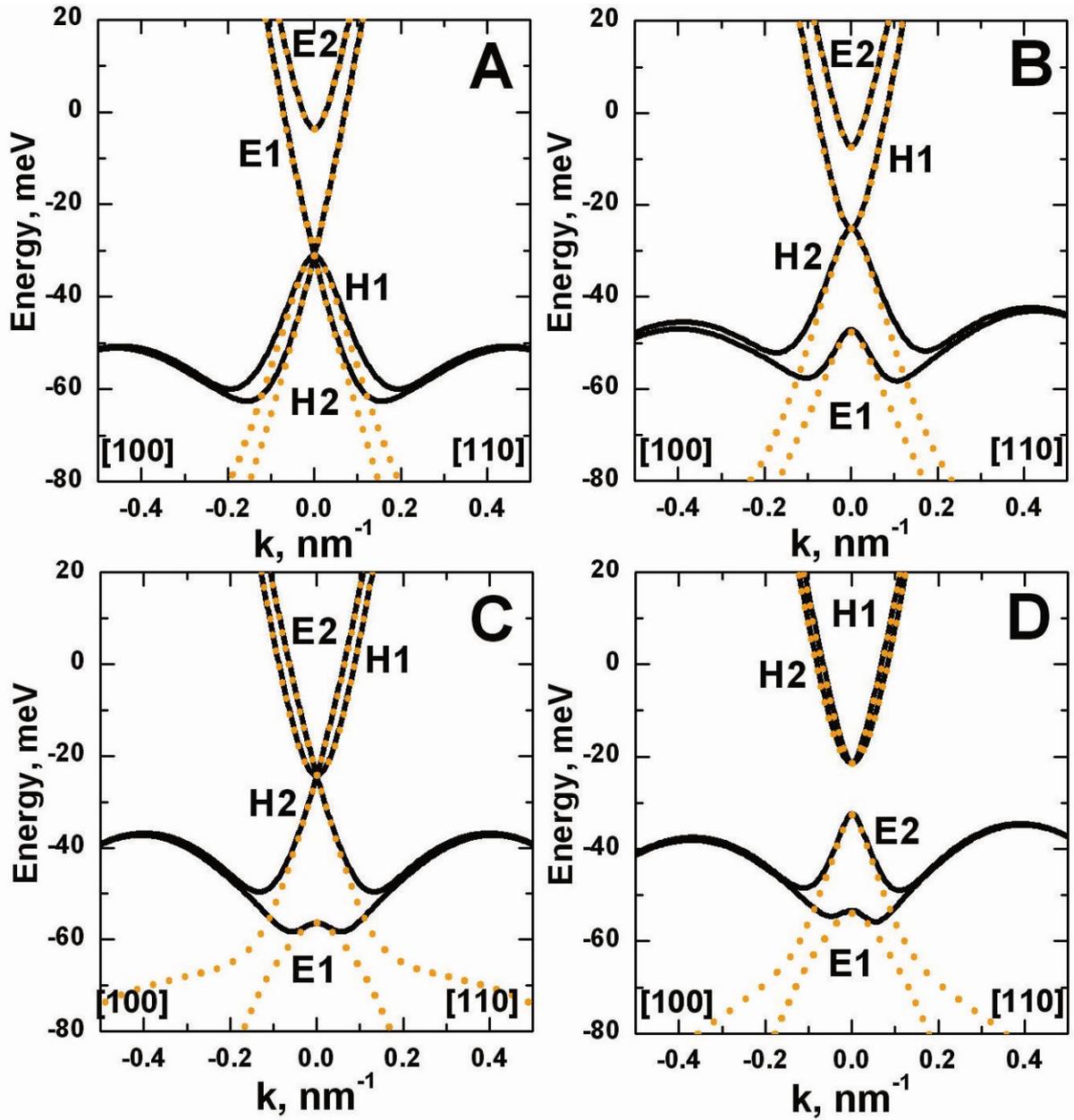

**Fig. S3**. (**A**) Comparison between calculations within the eight-band Kane model (black solid curves) and by using the effective Hamiltonian $H_{eff}(k_x, k_y)$ (orange dotted curves). The cases (**A**-**D**) are connected with those shown in Fig. 4.

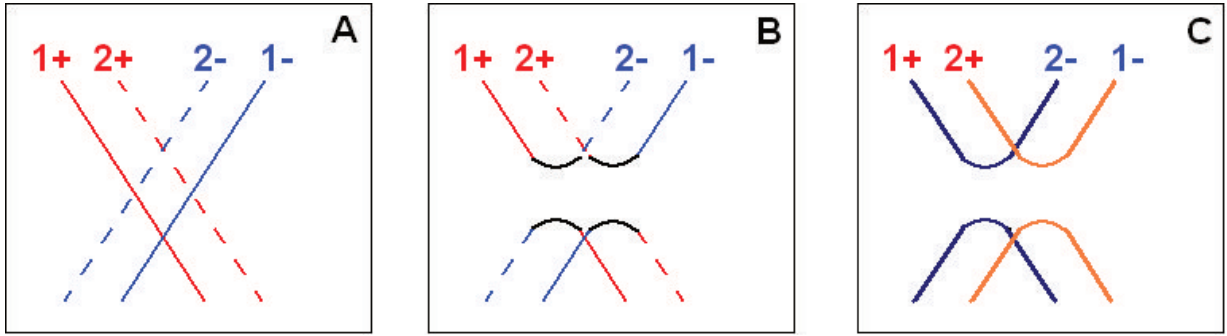

Fig. S4. Schematic formation of the edge states in two coupled 2D TI: (**A**) spin-conserved tunneling, (**B**,**C**) spin-dependent tunneling. The solid curves in the panels (A) and (B) present dispersion of the edge states in the first layer, while the dashed curves are the edge states in the second layer. Blue and red colors are for different spin orientation in each layer. The different colored curves in the panel (**C**) correspond to different Kramer's partners in the whole system.